\documentclass[11pt]{article}  
 \pdfoutput=1 
 \usepackage{jheppub}
 \usepackage{epsfig}
\usepackage{graphicx, color, xcolor}
\usepackage{amsmath,amsfonts,amssymb}
\usepackage{slashed}

\usepackage{tcolorbox}

\def\eps{\epsilon}

\def\al#1{\alpha_{#1}}
\def\eq#1{(\ref{#1})}
\newcommand{\beq}{\begin{equation}}
\newcommand{\eeq}{\end{equation}}
\newcommand{\bea}{\begin{eqnarray}}
\newcommand{\eea}{\end{eqnarray}}
\newcommand{\be}{\begin{equation}}
\newcommand{\ee}{\end{equation}}

\newcommand{\DDslash}[1]{\hspace*{0.1cm} \slash \hspace*{-0.26cm}{#1} }

\title{Safe  Gauge-String Correspondence}
\author{Soo-Jong Rey$^{1,2,3}$,}\author{Francesco Sannino$^{4}$}
\affiliation{$^{1}$ Institute for Quantum Advanced Research Korea (i-{\rm QUARK}), Seoul 06642 {\rm KOREA} \\
$^{2}$ School of Physics \& Astronomy, Seoul National University, Seoul 08220 {\rm KOREA} \\
$^{3}$ Fields, Gravity \& Strings, Institute for Basic Science, Daejeon 34126 {\rm KOREA}}   
 \affiliation{$^{4}$CP$^3$-Origins and D-IAS, Univ. of Southern Denmark, Campusvej 55, Odense M, 5023 \rm DENMARK}
  \abstract{Safe theories are quantum field theories whose continuum limit is defined by a non-Gaussian ultraviolet fixed point when the ultraviolet cutoff is removed. They constitute an important set in the space of quantum field theories. Here we develop the `safe' gauge-string correspondence program according to which $d$-dimensional safe gauge theories, admitting the `t Hooft-Veneziano limit, are holographically dual to $(d+1)$-dimensional safe noncritical string theories on asymptotically anti-de Sitter space. We provide evidences for this correspondence on a class of safe templates that engage fermion and scalar matter fields into gauge, Yukawa and Higgs self-interactions. Safe theories can feature in the infrared both a weak coupling phase and a strong coupling phase on either side of the ultraviolet fixed point. They correspond respectively to dilaton and warp factors taking domain-wall or Liouville-wall profiles on asymptotically anti-de Sitter space. We argue that four-dimensional ${\cal N}=4$ super Yang-Mills theories and all known interacting (super)conformal field theories are nonperturbative limit situations of safe gauge theories. The weak coupling phase  provides a solvable holographic renormalization group flow while the strong coupling infrared phase provides a runaway-free alternative to holographic QCD.}

\begin{document}
 \maketitle

 \section{Introduction}
The anti-de Sitter / conformal field theory (AdS/CFT) correspondence \cite{Maldacena:1997re} asserts the existence of a weak-strong duality between conformal field theories in $d$-dimensional spacetime and string theory in $(d+1)$-dimensional anti-de Sitter spacetime. The correspondence have been explored for various conformal field theories up to $d = 6$ dimensions, predominantly with supersymmetry to attain better analytic control. Among the most studied models, is ${\cal N}=4$ super Yang-Mills theory \cite{Brink:1976bc} in four dimensions which is dual to the ${\cal N}=4$ anti-de Sitter string theory in five dimensions. The ${\cal N}=4$ super Yang-Mills theory is known as the simplest quantum field theory. With the beta function vanishing exactly \cite{Brink:1982wv, Mandelstam:1982cb, Sohnius:1981sn, Novikov:1983uc,Howe:1983sr}), this theory is conformally invariant and its gauge coupling parameter takes an arbitrary value. An outstanding question is whether the AdS/CFT correspondence extends to the more intricate class of gauge and string theories.  For example, the Standard Model~\cite{Weinberg:1967tq, Salam:1968rm} of the real world\footnote{Authors are confident that string theorists write this statement and read it as a certain stack of D-brane complexes on an orbifold.} consisting of chiral fermions and Higgs scalar interact through chiral gauge, Yukawa and Higgs couplings and these couplings all run with energy scales. With three generations, the strong and weak sectors become asymptotically free in the ultraviolet while the hypercharge and Higgs sectors hit Landau poles toward the ultraviolet. 
 
It is widely viewed that the ${\cal N}=4$ super Yang-Mills theory, which describes the low-energy dynamics (below the string scale $M_{\rm s}$) of stacked D3-branes, is a limit situation\footnote{It would be interesting if the readers interpret this in the sense of Grenzsituation of Karl Jaspers.} of quantum chromodynamics (QCD) \cite{Fritzsch:1973pi}, the strong sector of the Standard Model. Modulo differences that the gauge group $SU(3)$ is enlarged to $SU(N_c)$ and the quarks transforming as the fundamental representation of $SU(3)$ are now fermions and scalars transforming as the adjoint representation of $SU(N_c)$ (these matter fields combine with gluons to form the ${\cal N}=4$ supermultiplet), in essence, both theories are vectorial and non-abelian gauge theories. As such, one might hope to better understand nonperturbative aspects of the QCD from the simpler, ${\cal N}=4$ super Yang-Mills theory by itself and from its AdS/CFT correspondence.

Variants of this view were further put forward by coupling the ${\cal N}=4$ super Yang-Mills theory (starting with D3-branes) to matter fields in fundamental representation (which amounts to introducing D7 and anti-D7-branes) \cite{Karch:2002sh} or by embedding the ${\cal N}=4$ super Yang-Mills theory into five dimensions (now starting with D4-branes), then compactified on a circle with thermal boundary conditions \cite{Witten:1998zw} and also by further adding quarks in the fundamental and anti-fundamental representations (which amounts to introducing D8 and anti-D8 branes) \cite{Sakai:2004cn}. It was found that these setups captured numerous low-energy manifestations of  non-abelian Yang-Mills theories and, in particular of QCD, yet because of the dramatic differences in their ultraviolet completion, they are left unsolved with multitudes of technical and conceptual difficulties and puzzles. 

For instance, what sets the ${\cal N}=4$ super Yang-Mills dynamics apart from the asymptotically free QCD \cite{Gross:1973ju,Politzer:1973fx} is that its coupling is fixed and finite at all scales. Because of this conformal invariance, it is impossible to develop a mass gap in the infrared (even if the couplings are sent to infinity). While the conformal invariance may be softly broken by fermion and scalar masses (and decouple them from the infrared dynamics), it is also impossible to develop partons exhibiting the Bjorken scaling in the ultraviolet (unless the coupling is tuned to zero)\footnote{We recall that the notion of parton is well defined only when the quantum field theory is non-interacting or super-renormalizable. Whether partons are well defined in the  ${\cal N}=4$ super Yang-Mills theory at strong coupling was explored in \cite{Polchinski:2002jw}  by utilizing the AdS/CFT correspondence. The answer is, as expected, negative.}. 
This casts doubts to the aforementioned connection and poses an interesting question as to what class of gauge theories the ${\cal N}=4$ super Yang-Mills theory can be more closely be seen as a limit situation. 

In this paper, we assert that the ${\cal N}=4$ super Yang-Mills theory is better interpreted as a limit situation of safe gauge theories than of asymptotically free ones. By the same token, we also assert that all known interacting conformal field theories in various dimensions and with various amount of supersymmetries, are also better interpreted as a limit situation of safe gauge theories than of asymptotically free ones. 
Moreover, we develop a {\it safe} gauge-string correspondence for a class of safe gauge theories\footnote{A natural limit  of the {\it safe} gauge-string correspondence is the conventional AdS/CFT correspondence.}. By safe gauge theory, we refer to interacting gauge theory whose existence in the infinite cutoff limit is defined by a non-Gaussian ultraviolet stable fixed point. The theory may flow out toward the infrared to a weak coupling phase of either Gaussian (like QED) or non-Gaussian (viz. interacting) fixed point. Alternately the theory may flow toward the infrared to a strong coupling phase of color confinement and dynamical chiral symmetry breaking. According to our assertion the ${\cal N}=4$ super Yang-Mills theory is the simplest, featureless situation that starting from an ultraviolet fixed point the theory does not flow to the infrared. Stating our assertion differently, safe gauge theories better match their dynamics with the dynamics of ${\cal N}=4$ super Yang-Mills theory\footnote{Another simplest safe (gauge) theory is the free field theory. Its beta function is trivially zero, and its coupling vanishes. It is the theory defined by an isolated Gaussian fixed point. In all known cases, a safe gauge theory would flow between two interacting fixed points or between an ultraviolet interacting fixed point and an infrared QCD-like, mass-gapped dynamics.}. 

\begin{figure}[htb]
\centering
\includegraphics[width=.6\textwidth, height=0.2\textheight]{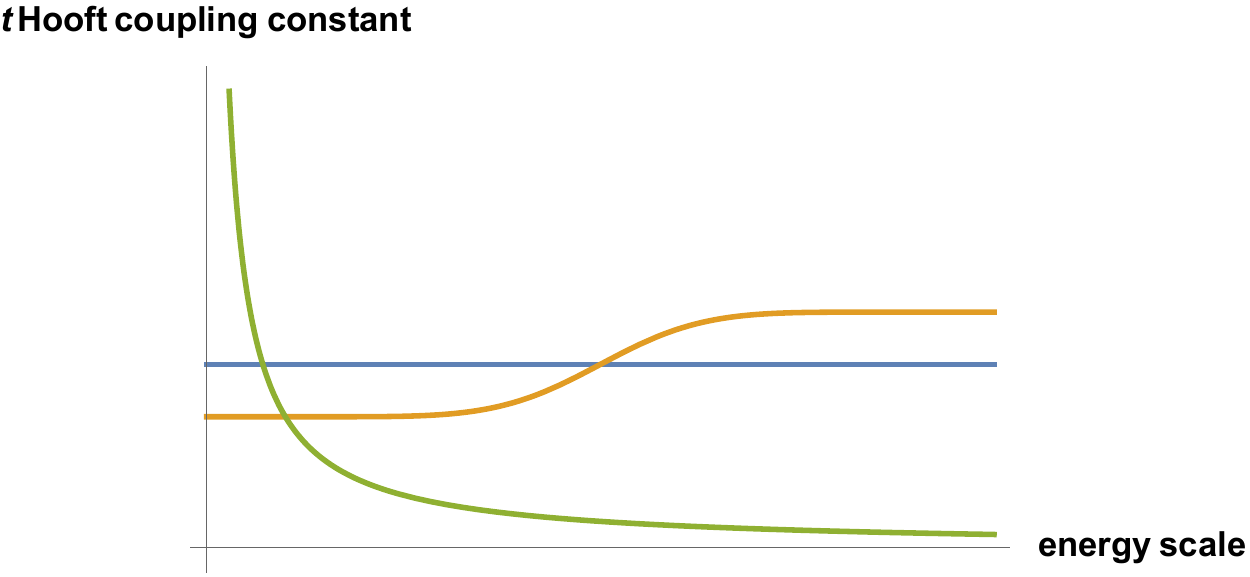}
\caption{\label{RCmassless} 
Coupling constant running is illustrated for ${\cal N}=4$ super Yang-Mills theory (blue), safe gauge theory (orange), and asymptotically free gauge theory (green).  The ultraviolet fixed point is non-Gaussian for safe theories and Gaussian for asymptotically free theories, respectively. It displays that the ${\cal N}=4$ super Yang-Mills theory is more a limit situation of the safe gauge theories than of the asymptotically free gauge theories. The scales are arbitrary.}
\end{figure}

Intuitively, we can motivate our assertion by comparing characteristic running of gauge coupling constants in asymptotically free gauge theory, ${\cal N}=4$ super Yang-Mills theory (as a representative of interacting conformal field theories), and safe gauge theory, respectively. When all matter fields (by these, we mean fields other than the gauge fields) are massless, the gauge coupling constant of each theory runs from asymptotic zero to infinity toward the infrared, sits at a fixed value, or interpolates from an interacting ultraviolet fixed point to an infrared fixed point, respectively. Their behavior, illustrated in Fig.\ref{RCmassless}, suggests that the ${\cal N}=4$ super Yang-Mills theory would dynamically behave more like safe gauge theories than asymptotically free gauge theories.  

\begin{figure}[htb]
\centering
\includegraphics[width=.6\textwidth]{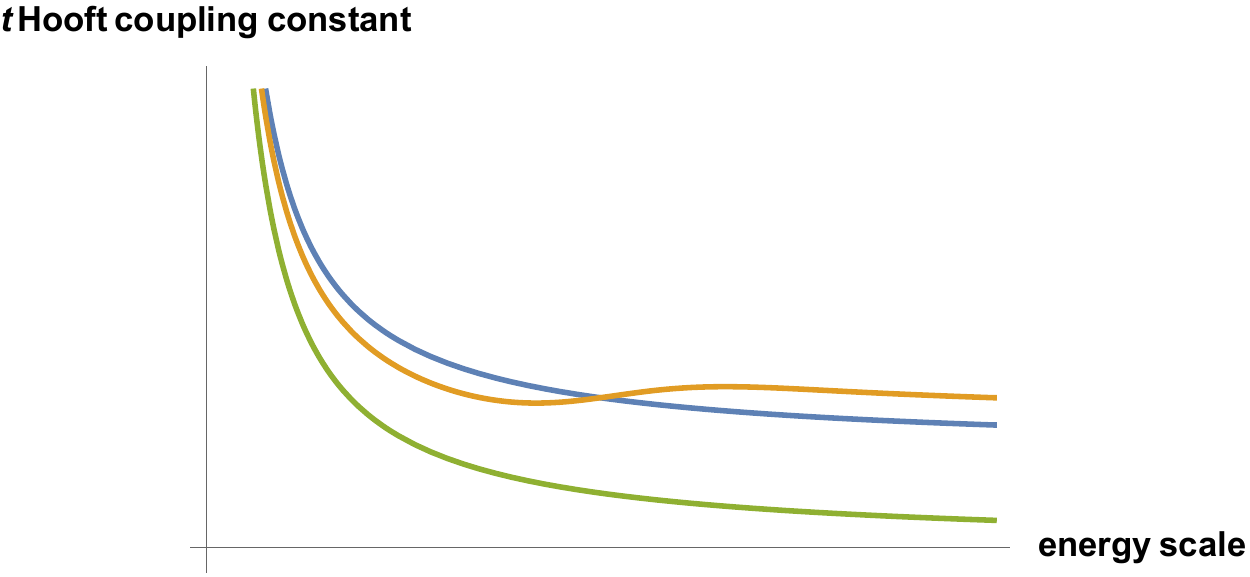}
\caption{\label{RCmassive} Coupling constant running is illustrated for  ${\cal N}=4$ super Yang-Mills theory (blue), safe gauge theory (orange) and asymptotically free gauge theory (green), after a mass deformation is introduced at a common scale. It again shows that the ${\cal N}=4$ super Yang-Mills theory is a limit situation of the safe gauge theories rather than of the asymptotically free gauge theories. The scales are arbitrary.}
\end{figure}

The similarity extends if the matter fields are made massive. Below the mass scale, all three theories behave effectively as pure Yang-Mills theory and their coupling constants runs to infinity toward the scale of dynamically generated mass gap. Above the mass scale, the running coupling constant of respective theory approaches the massless counterpart in the ultraviolet. Their behavior, illustrated in Figure~\ref{RCmassive}, again points to the suggestion that the ${\cal N}=4$ super Yang-Mills theory behaves more like safe gauge theories than asymptotically free gauge theories. 

Turning our proposal around,  the ${\cal N}=4$ super Yang-Mills theory furnishes non-perturbative data points in the space of safe gauge theories, as its ultraviolet fixed point is exact and forms a line of stable fixed points (the $\mathbb{C}^1$ conformal manifold) to an arbitrarily strong coupling. The same also holds for all known interacting conformal field theories in various dimensions and with various supersymmetries, either isolated at a fixed coupling or varying over a conformal manifold. By the AdS/CFT correspondence, these theories in the strong coupling regime are most effectively described in terms of the anti-de Sitter string theories in the weak curvature regime. With this paper, we extend this holographic description to more structured safe gauge theories. In four dimensions, a class of safe gauge theories was constructed in \cite{Litim:2014uca,Litim:2015iea}. These theories are akin to the Standard Model and involve gauge, Yukawa and scalar couplings and interpolate between a free or interacting fixed point in the infrared and an interacting fixed point in the ultraviolet. Taking them as a template, we develop the `safe' gauge-string correspondence. 

\begin{figure}[htb]
\centering
\includegraphics[width=.3\textwidth]{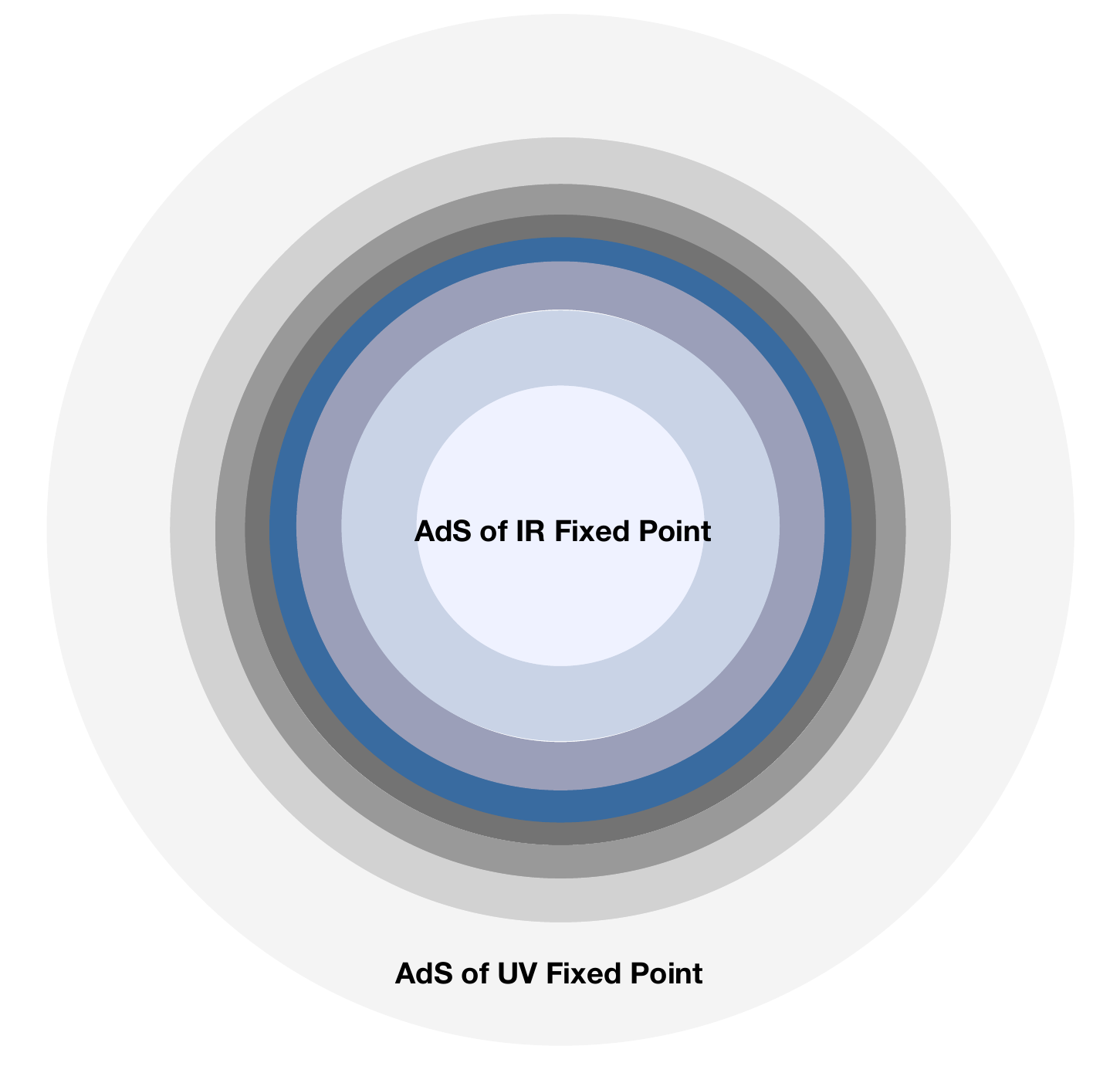} \hskip .6cm
\includegraphics[width=.5\textwidth]{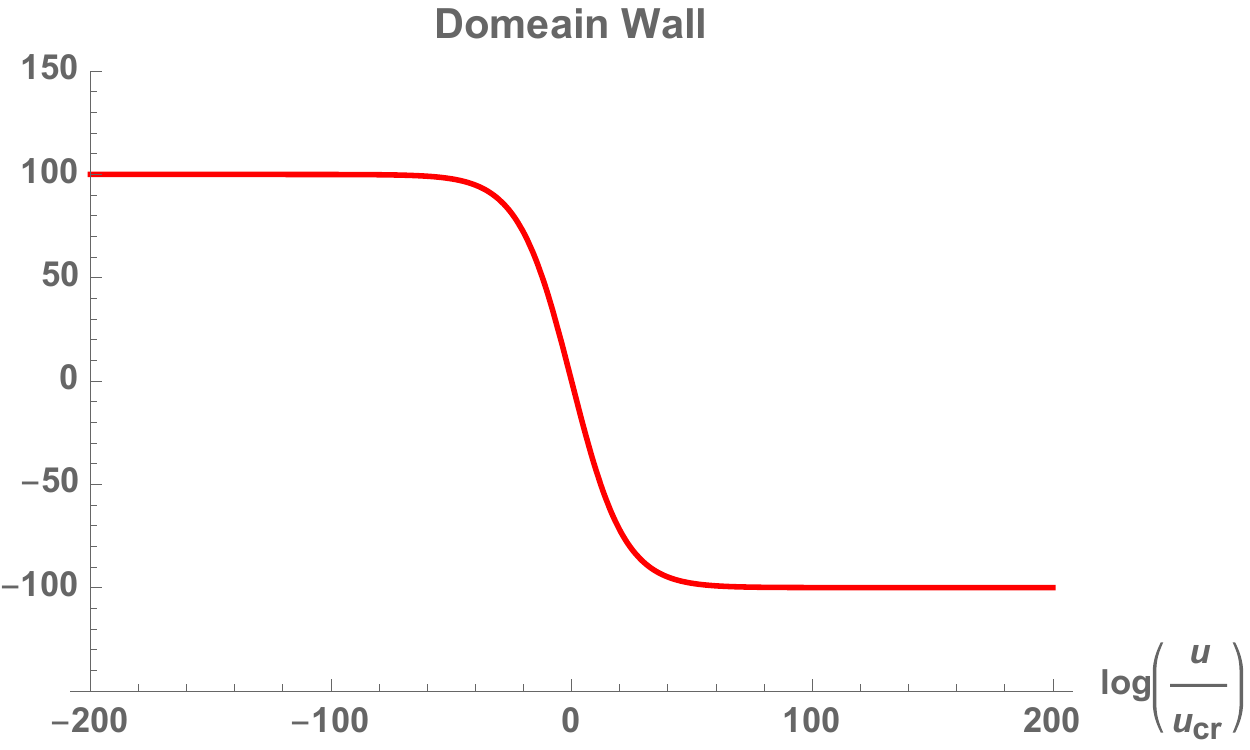}
\caption{\label{domainwall} The dilaton and warp factor in the weak coupling phase of the safe string theory are expected to have the the domain wall profile. They asymptote to the boundary values toward $u = 0$ and $u = \infty$ (along the abscissa in the log scale), respectively, and represents the coupling parameters at the ultraviolet and the infrared fixed points, respectively, along the line of physics of the safe gauge theory. In the Figure, $u_{\rm cr}$ is the critical position of the domain wall (holographic dual of the critical transition scale $1/\Lambda_{\rm cr}$). The scale of the ordinate is arbitrary.} \end{figure}

An important aspect of safe gauge theories is the existence of, at least, two possible infrared dynamics; relative to the ultraviolet fixed point, there can be a weak infrared coupling phase and a strong interacting coupling phase, respectively. While the weak coupling phase is analytically more amenable, the strong coupling phase is also of equal interest, as it exhibits essentially the same infrared dynamics as the QCD and so rightfully referred to as {\it safe QCD}. In this paper, we will study both phases; accordingly, we will show that the holographic dual of this safe template is given by either domain wall (as illustrated in Figure~\ref{domainwall}) or Liouville (confining) wall background in asymptotically anti-de Sitter space. 

We organize this paper as follows. In Section \ref{Safe}, we review a class of safe gauge theories constructed in \cite{Litim:2014uca}, which we refer to as the safe template, and collect its salient features to be compared with those of the ${\cal N}=4$ super Yang-Mills theory. The idea is to bring to light as many similarities between these two theories as possible and to lend support of our assertion that the ${\cal N}=4$ super Yang-Mills theory is a time honored, limit of safe gauge theories. We do so first for the weak coupling phase mainly as it is amenable within analytically controlled perturbative methods, and then for the strong coupling phase. We also summarize known results for supersymmetric counterparts and relegate detailed analysis to separate works.  These observations allow us, in Section~\ref{label}, to develop the safe gauge-string correspondence that, as for the ${\cal N}=4$ super Yang-Mills theory, safe gauge theories admit holographic duals for both weak and strong coupling phases. Built upon this result, we propose that safe $d$-dimensional quantum field theories admitting `t Hooft  \cite{tHooft:1974pnl} or 't Hooft -Veneziano \cite{Veneziano:1976wm} limits are holographically dual to $(d+1)$-dimensional asymptotically anti-de Sitter string theories with domain wall or Liouville wall deformations for both the dilaton and the warp factor~\footnote{The Veneziano limit in holographic QCD was explored in \cite{Jarvinen:2011qe}. They addressed different questions from ours.}. In Section \ref{Conclusions}, we summarize our proposal and discuss various future directions to advance the proposed program.

\section{The Safe Gauge Theory}
\label{Safe}
In this section, we present a template of the safe gauge theory and discuss its variants. The template is the gauge-Yukawa-scalar theory, first constructed in \cite{Litim:2014uca}. We highlight salient aspects of the template and note that they strikingly resemble characteristics of the ${\cal N}=4$ super Yang-Mills theory. We claim that these features are universal to all safe gauge theories in four dimensions and that the ${\cal N}=4$ super Yang-Mills theory is nothing but a subclass of the safe gauge theories at strong couplings. In the next section, we will draw support to our claims from holographic dual description by establishing the gauge-string correspondence.

\subsection{Safe Template}
\label{SafeT}
The safe template of \cite{Litim:2014uca} contains $(N_c^2-1)$ non-Abelian gauge fields, $A_\mu^i$, $N_c N_f$ massless Dirac fermions, $\Psi = (\psi_L, \psi_R)$, and $N_f^2$ massless complex-valued scalars, $H$. The theory is symmetric under the local $G_g = SU(N_c)$ gauge transformation and the classical global $U(N_f)_L \times U(N_f)_R$ flavor transformations. At quantum level, the Adler-Bell-Jackiw axial anomaly reduces the classical global symmetry to the quantum one, $G_f = SU(N_f)_L \times SU(N_f)_R \times U(1)_V$. The left- and right-handed Weyl fermions transform, respectively, in $(N_f,0)$ and $(0,N_f)$ representations of the flavor symmetry group $G_f$: 
\begin{equation}
\psi_L \longrightarrow \psi_L^\prime = U_L \, \psi_L\;,\qquad 
\psi_R \longrightarrow \psi_R^\prime = U_R \, \psi_R\;,
\end{equation}
while the scalar field $H$ transforms in $(N_f,\overline{N}_f)$ bifundamental representation of the flavor group $G_f$:
\begin{equation}
H \longrightarrow H^\prime = U_L \, H \, U_R^\dagger\;
\end{equation}
for $U_{L, R}$ the rotations in $SU(N_f)_{L, R}$. 
We refer to this set of fermion and scalar field as the minimal contents. One may introduce multiple (but finite) copies of these matter fields. In that case, a residual global symmetry $G_r$ may further arise that rotates the matter field multiplets among themselves. Hereafter, we limit ourselves to the minimal contents but we will keep the option in mind in foregoing discussions. 

The Lagrangian reads, in canonical normalization of kinetic terms, 
\begin{eqnarray} \label{Lag}
{\cal L} &=&  - \frac{1}{4}\, \left( F_{\mu \nu}^i F^{\mu \nu \, i} \right) + \bar{\psi}_L\, i \DDslash{D} \, \psi_L + i \bar{\psi}_R\, i \DDslash{D} \, \psi_R
+ {\rm Tr} \left( \partial_\mu H^\dagger\, \partial^\mu H \right)  \nonumber \\
& +& y \left( \bar{\psi}_L \, H \, \psi_R + \bar{\psi}_R \, H^\dagger \psi_L \right)
- u \, {\rm Tr} \left( H^\dagger H \right)^2 - v \left[ {\rm Tr} \left( H^\dagger H \right) \right]^2\;,
\end{eqnarray}
with the covariant derivative
\begin{equation}
D_\mu = \partial_\mu + ig A_\mu^i t_i\;.
\end{equation}
Here, $g$ is the gauge coupling constant of the non-Abelian gauge sector and $t_i, \, i = 1, \ldots,
N_c^2-1$, are the generators of $SU(N_c)$ in the fundamental representation. Tr refers to the trace over the $G_f$ global symmetry. We define the safe theory as the one with exact $G_f$ flavor symmetry (which includes single and double trace interactions) and so do not allow any deformations that breaks $G_f$ flavor symmetry. As such, the four marginal couplings $(g, y, u, v)$ are all there are for spanning the theory space and they are all singlets under the $G_f$. 

The theory is spanned by, in addition to $N_c$ and $N_f$, the aforementioned four marginal couplings: the gauge coupling $g$, the Yukawa coupling $y$, the quartic scalar coupling ${u}$, and the double-trace scalar coupling $v$. Altogether, there are six parameters that are needed for specifying the safe theory \footnote{For brevity, we omit the flavor-diagonal CP violating $\theta$ angle parameter. It is just the real part of the complex gauge coupling parameter.}. We will explore the theory in the Veneziano limit \footnote{We stress that the exact $G_f$ symmetry is imperative for defining the Veneziano limit unambiguously. In this regard, the theory space we defined by limiting to the four aforementioned marginal couplings is self-consistent.}:
\beq
 N_c \rightarrow \infty,  \qquad N_f \rightarrow \infty, \qquad {N_f \over N_c} = \mbox{fixed},
 \nonumber
 \eeq
for which the expansion parameters are $1/N_c, 1/N_f$ and the rescaled four marginal couplings
\beq\label{couplings}
\al g=\frac{g^2\,N_c}{(4\pi)^2}\,,\quad
\al y=\frac{y^{2}\,N_c}{(4\pi)^2}\,,\quad
\al h=\frac{{u}\,N_f}{(4\pi)^2}\,,\quad
\al v=\frac{{v}\,N^2_f}{(4\pi)^2}\,.
\eeq
The appropriate powers of $N_c$ and $N_f$  are to prepare for the  
the Veneziano limit. In the Veneziano limit, we also introduce the so-called the Veneziano ratio \footnote{In Feynman diagrammatics, this parameter defines a renormalized fugacity for inserting a flavor index loop in the fishnet of color loops.}
\begin{equation}\label{eps}
\epsilon=\frac{N_f}{N_c}-\frac{11}{2} = \mbox{fixed}
\end{equation}
as a continuous parameter. 
Thus, at each fixed value of $\epsilon$, we see that in the Veneziano limit the number of flavors $N_f$ is interlocked to the number of colors $N_c$. Accordingly, at each fixed value of $\epsilon$, the flavor symmetry $G_f$ is also interlocked to the gauge symmetry $G_c$. 
\begin{tcolorbox}
{\bf Safe~1}: In safe theories, the flavor global symmetry $G_f$ defined by fermion and scalar fields is interlocked to the gauge symmetry $G_c$. The residual global symmetry $G_r$ may remain independent.
\end{tcolorbox}

In the Veneziano limit, the topological loop expansion is set by $1/N_c^2$. For a fixed $N_c$, the remaining five coupling parameters are the parameter $\epsilon$ and the four marginal couplings. To further study them, we now examine the renormalization group flows of the theory. The relevant beta functions of the theory was obtained in Ref.\ \cite{Antipin:2013pya} in dimensional regularisation. We use the short-hand notation $\beta_i\equiv\partial_t\alpha_i$, with $i=(g,y,h,v)$ and $t$ the logarithm of energy scale, for the beta functions of the respective couplings \eq{couplings}. 

Restricted to the gauge sector, viz. in the subspace of the coupling $\alpha_g$, the theory is asymptotically free for $\eps < 0$, while it is infrared free for $\eps>0$ at which the asymptotic freedom is lost and the gauge dynamics behaves like non-abelian QED. Intriguingly, charted to the full sector, viz. in the full space of the four couplings $\alpha_g, \alpha_y, \alpha_h, \alpha_v$, it was discovered in \cite{Litim:2014uca} that the theory exhibits an interacting ultraviolet fixed point. It turns out that, provided $0 < \eps \ll 1$, this fixed point can be controlled within perturbation theory.

After a careful study of the zeros of the beta functions and taking account of the stability of the quantum potential \cite{Litim:2014uca,Litim:2015iea},  one arrives at the unique mathematically and physically acceptable fixed point accessible within perturbation theory:  
\beq\label{alphaNNLO}
\begin{array}{rcl}
\alpha_g^* \Big\vert_{\rm uv} &=&
+\frac{26}{57}\,\eps+ \frac{23 (75245 - 13068 \sqrt{23})}{370386}\,\eps^2
+{\cal O}(\eps^3)
\\[1.ex]
\alpha_y^*\Big\vert_{\rm uv} &=&
+\frac{4}{19}\,\eps+\left(\frac{43549}{20577} - \frac{2300 \sqrt{23}}{6859}\right)\,\eps^2
+{\cal O}(\eps^3)
\\[1ex]
\alpha_h^*\Big\vert_{\rm uv} &=&
+\frac{\sqrt{23}-1}{19}\,\eps+{\cal O}(\eps^2)
\,,\\[1.ex]
\al {v}^*\Big\vert_{\rm uv} &=&
-\frac{1}{19} \left(2 \sqrt{23}-\sqrt{20 + 6 \sqrt{23}}\right)\,\eps
+{\cal O}(\eps^2)\,.
\end{array}
\eeq
We expressed the fixed point in the analytic expansion of the parameter $\epsilon$. We see that, remarkably, the parameter $\epsilon$ is interlaced to the fixed point values of the four coupling parameters. We expect the latter to be a robust feature of the safe theory.
\begin{tcolorbox}
{\bf Safe~2}: In safe theories, the ratio $N_f/N_c$ is intertwined to the values of the couplings at the non-Gaussian ultraviolet fixed point.
\end{tcolorbox} 

The existence of such fixed point ensures that the theory is ultraviolet complete, and so the theory is valid at arbitrarily short distance scales. In particular,  scalar interactions are indispensable while remaining free of the triviality problem because of the presence of an interacting ultraviolet fixed point. This is a feature shared by the scalar interactions in the ${\cal N}=4$ super Yang-Mills theory, but attained without the help of supersymmetry. 
\begin{tcolorbox}
{\bf Safe~3}: In safe theories, the scalar fields self-interact while avoiding the triviality problem. Moreover, this property holds without supersymmetry. 
\end{tcolorbox}

The template is also infrared free, since at low energies it is governed by the Gaussian fixed point:
\beq
\alpha_g^* \Big\vert_{\rm ir} = \alpha^*_y  \Big\vert_{\rm ir}= \alpha_h^*  \Big\vert_{\rm ir}= \alpha_v^*  \Big\vert_{\rm ir} = 0.
\nonumber
\eeq
 The renormalization group flow connecting the Gaussian infrared fixed point and the non-Gaussian ultraviolet fixed point is the ultraviolet-complete trajectory, characterising a line of physics. The latter is also known as {\it separatrix}, as it separates different physical regions of the theory in the space of renormalization group flows. Along the line of physics, the theory is non-interacting in the infrared and hits the above interacting fixed point in the ultraviolet. The region of the renormalization group phase diagram emanating from the ultraviolet-stable fixed point and leading to stable trajectories is known as the {\it ultraviolet critical surface}. Remarkably this critical surface is 
one-dimensional \cite{Litim:2014uca} and it has, clearly, a dynamical nature. This is so in every known safe theories, and so we promote it as a universal feature\footnote{We are not  considering theories in which some of the couplings become free in the ultraviolet and, in any event, we can always consider safe theories with one dimensional critical surfaces.}. 
\begin{tcolorbox}
{\bf Safe~4}: In safe theories, the couplings flow in scale over a finite but arbitrary interval set by $N_f / N_c$. We focus on proper safe theories in which the critical surface is one-dimensional.
\end{tcolorbox}

As elaborated above, the separatrix connects the non-Gaussian ultraviolet fixed point with the Gaussian infrared fixed point and it coincides with the critical surface near the ultraviolet fixed point \cite{Litim:2014uca}. Anticipating a holographic dual description, it is illuminating to consider an analytical approximation that is accurate in the limit of vanishing $\epsilon$, yielding the following relations along the separatrix \cite{Litim:2015iea}:
\beq
\label{critsurf}
\begin{array}{rcl}
\al y &=&
 \frac{6}{13}\, \al g \ , \\[2ex]
\al h & = &
\frac{3}{26} \left(\sqrt{23}-1\right)\,\al g \ , \\[2ex]
\al v & = &
\frac{3}{26} \left(\sqrt{20 + 6\sqrt{23}}-2\sqrt{23}\right)\,\al g \,.
\end{array}
\eeq
The one-dimensional nature of the line of physics is encoded into the fact that in order to determine the running of all three other couplings it is sufficient to specify the running of the gauge coupling. Moreover, according to {\bf Safe~4}, the Veneziano ratio $\epsilon$ is completely subsumed in the specification of the critical surface, so the separatrix is the complete specification of the line of physics at each order in the $1/N^2_c$ expansion. 

One can determine the precise analytic running of the gauge coupling Ref.\ \cite{Litim:2015iea}; at running mass scale $\mu$, it reads 
\beq\label{critg}
\alpha_g(\mu) =
\frac{\alpha_g^\ast}{1+W(\mu) }
\ ,
\eeq
where $W(\mu)\equiv W[z(\mu)]$ is the Lambert function (also known as product logarithm function) defined by the relation 
$W(z) \exp W(z) = z$ with 
\bea
\label{W}
W&=&\frac{\alpha_g^*}{\alpha_g}-1\qquad {\rm and} \qquad
z= \left(\frac{\mu_0}{\mu}\right)^{\frac{4\epsilon}{3}\alpha_g^*}
\left(\frac{\alpha_g^*}{\alpha_g^0}-1\right)
\exp\left(\frac{\alpha_g^*}{\alpha_g^0}-1\right)\,.
\eea
Here, $\alpha_g^0$ is the gauge coupling at an arbitrary reference scale $\mu_0$, with $\mu/\mu_0$ 
ranging between $0$ and $\infty$ and the gauge coupling ranging between 
$0<\alpha_g^0<\alpha_g^\ast$.  Inserting Eq.\ \eq{critg} into Eq.\ \eq{critsurf}, we obtain an analytic description 
of the renormalization group evolution of  all couplings along the separatrix.  

At asymptotically high energies $W(\mu)$ vanishes; towards the infrared, it grows. It is convenient to fix $\alpha^0_g$ as a sequence of fraction
$\alpha^0_g \equiv \alpha^*_g/(1+k)$ with $k \in \mathbb{R}_+$, 
which in practice amounts to fixing the arbitrary renormalization reference scale $\mu_0$ along the renormalization group flow. As pointed out in Ref.\ \cite{Litim:2015iea},  the value $k=1/2$, viz. $\alpha^0_g = 2\alpha^*_g/3$, 
corresponds to the exact critical transition scale $\mu_0 = \Lambda_{\rm cr}$ above which the physics is dominated by the ultraviolet fixed point and below which it is governed by the infrared fixed point.  
\begin{tcolorbox}
{\bf Safe~5}: In safe theories, the gauge coupling interpolates sharply between the infrared fixed point behavior and the ultraviolet fixed point behavior across the exact critical transition scale $\Lambda_{\rm cr}$.
\end{tcolorbox}
The interacting nature of the ultraviolet fixed point is expressed by the behavior
that it is approached by a power-law scaling in the renormalization mass scale 
\begin{equation}
\alpha_g (\mu ) \simeq \alpha^\ast_g + (\alpha_g^0 - \alpha_g^*) \left( \frac{\mu}{\overline{\mu}_0}
\right)^{-\tfrac{104}{171} \epsilon^2} \, ,
\label{runningcoupling}
\end{equation}
where  $\overline{\mu}_0 =  \mu_0 (1 + \cal{O}(\epsilon))$. Here, we used Eq.\ (\ref{alphaNNLO}) and that, in the asymptotic ultraviolet limit, the Lambert function approaches zero as
\begin{equation}
W(\mu) \quad \propto  \quad
\left( \frac{\mu}{\mu_0}\right)^{-\tfrac{104}{171}\epsilon^2}
\qquad \mbox{as} \qquad {\mu \over \mu_0} \to \infty \, . 
\end{equation}
\begin{tcolorbox}
{\bf Safe 6}: In safe theories, the scaling violation near the ultraviolet fixed point exhibits power-law behavior, approaching the fixed point faster than in the asymptotically free case. 
\end{tcolorbox}
\noindent This power-law scaling violation (which is on top of the running of the coupling constant) was previously known in the context of parton model description of the deeply inelastic scattering \cite{Kogut:1973ub}. We refer to Eq.(\ref{runningcoupling}) as Kogut-Susskind scaling. 
\begin{figure}[htb]
\begin{center}
\includegraphics[width=.46\textwidth]{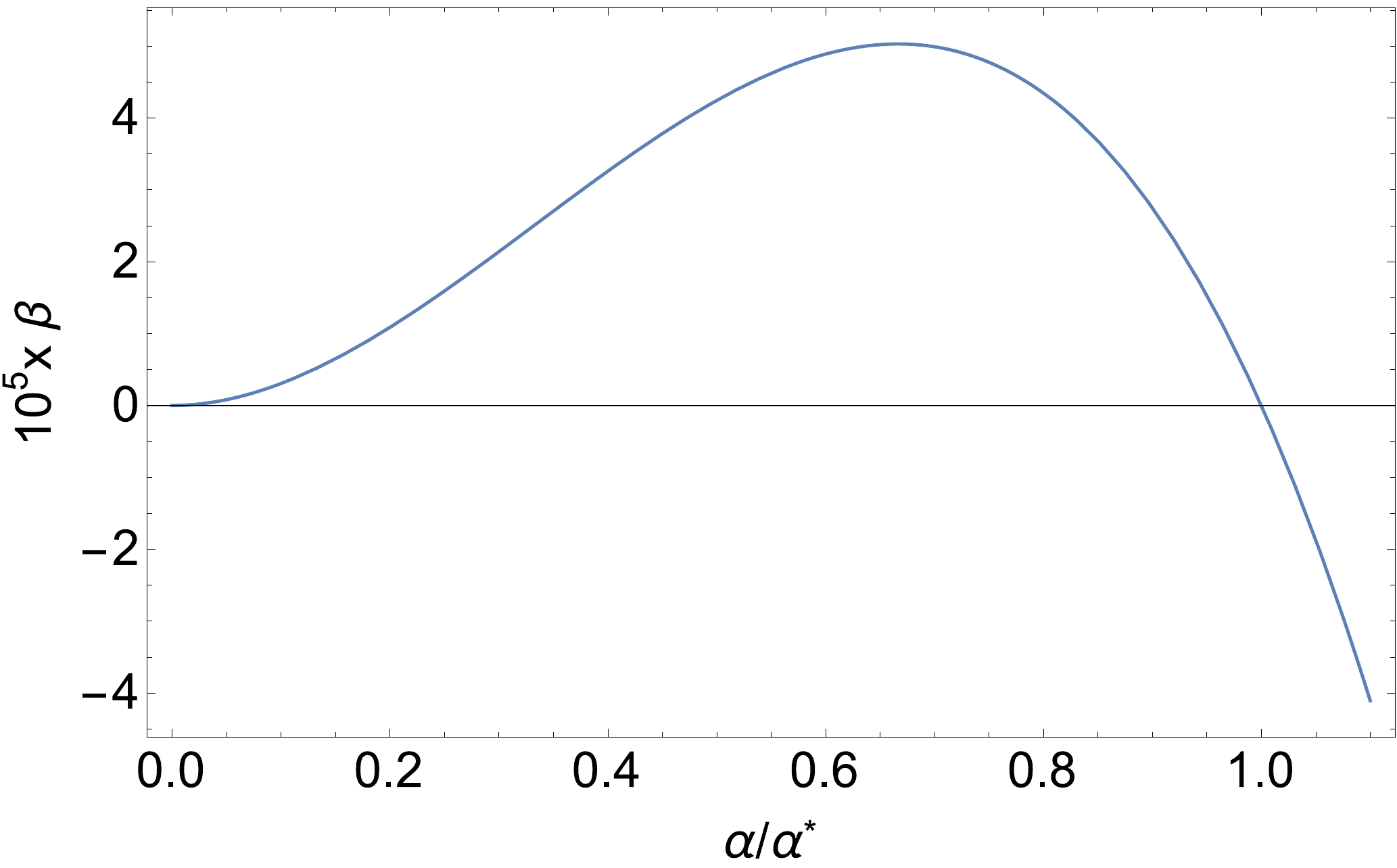} \hskip .4cm
\includegraphics[width=.46\textwidth]{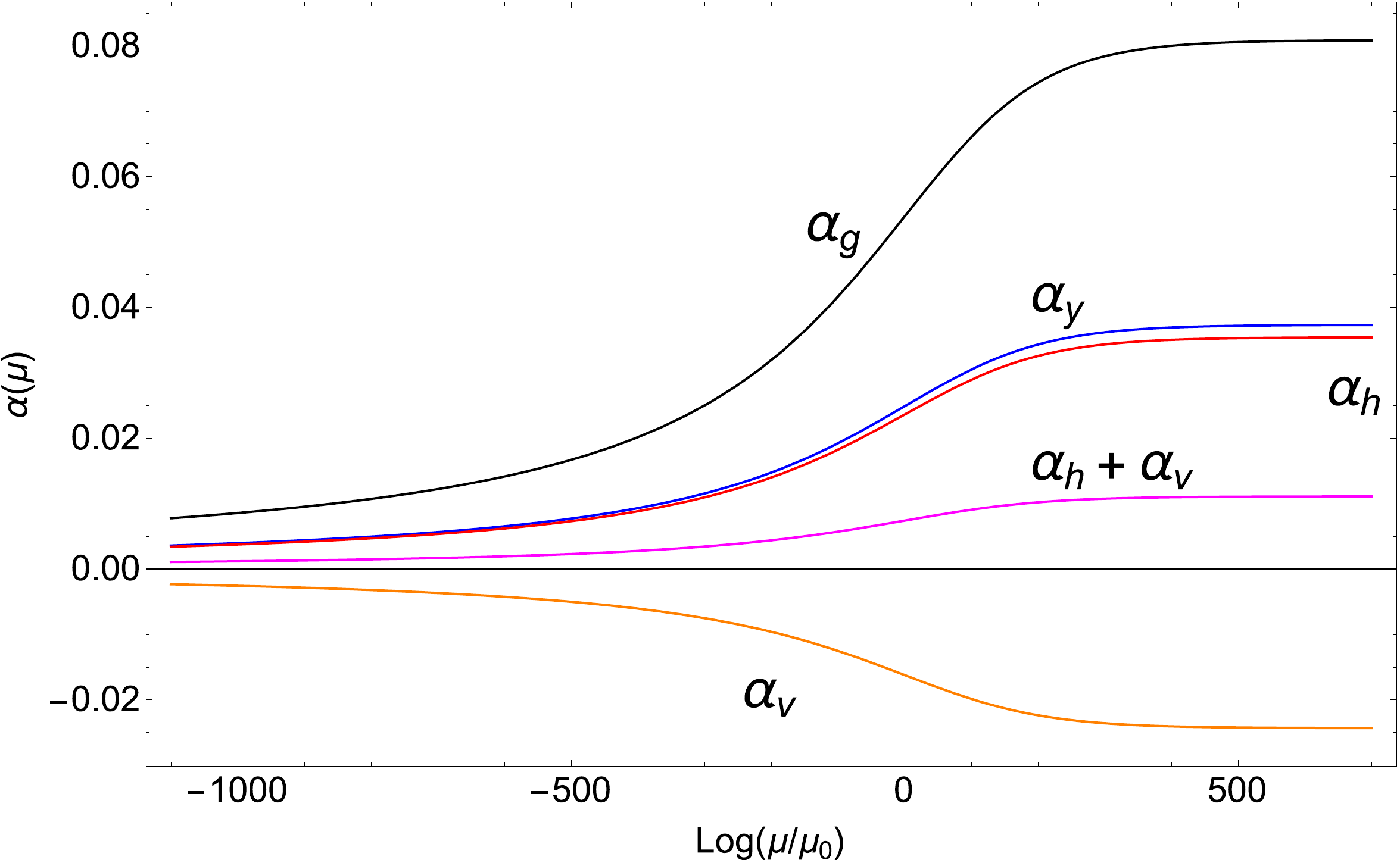} 
\caption{Left panel: Gauge beta function along the line of physics displaying the characteristic 
asymptotically safe behavior. Right panel: From top to bottom the running of the gauge,
Yukawa, double trace, single trace plus double trace, and only single trace coupling is shown
along the line of physics.  We have chosen $k=1/2$ and $\epsilon=0.1$.}
\label{running}
\end{center}
\end{figure}
In the left panel of Fig.\ \ref{running}, we display the beta function of the 
gauge coupling along the separatrix. In the right panel of Fig.\ \ref{running}, using Eqs.\ \eqref{critsurf} and \eqref{critg}, we also display the running of the four couplings along the separatrix. The {\it complete} asymptotically safe nature of the theory should now be clear: each of the four coupling parameters approach the nontrivial fixed point value toward the ultraviolet. We stress that the scalar sector, through its Yukawa interactions, was indispensable for the existence of the non-Gaussian ultraviolet fixed point, at least within the perturbative regime we examined. 

\subsection{Safe Template II: Safe QCD}
\label{2.2}
\begin{figure}[htb]
\centering
\includegraphics[width=.8\textwidth]{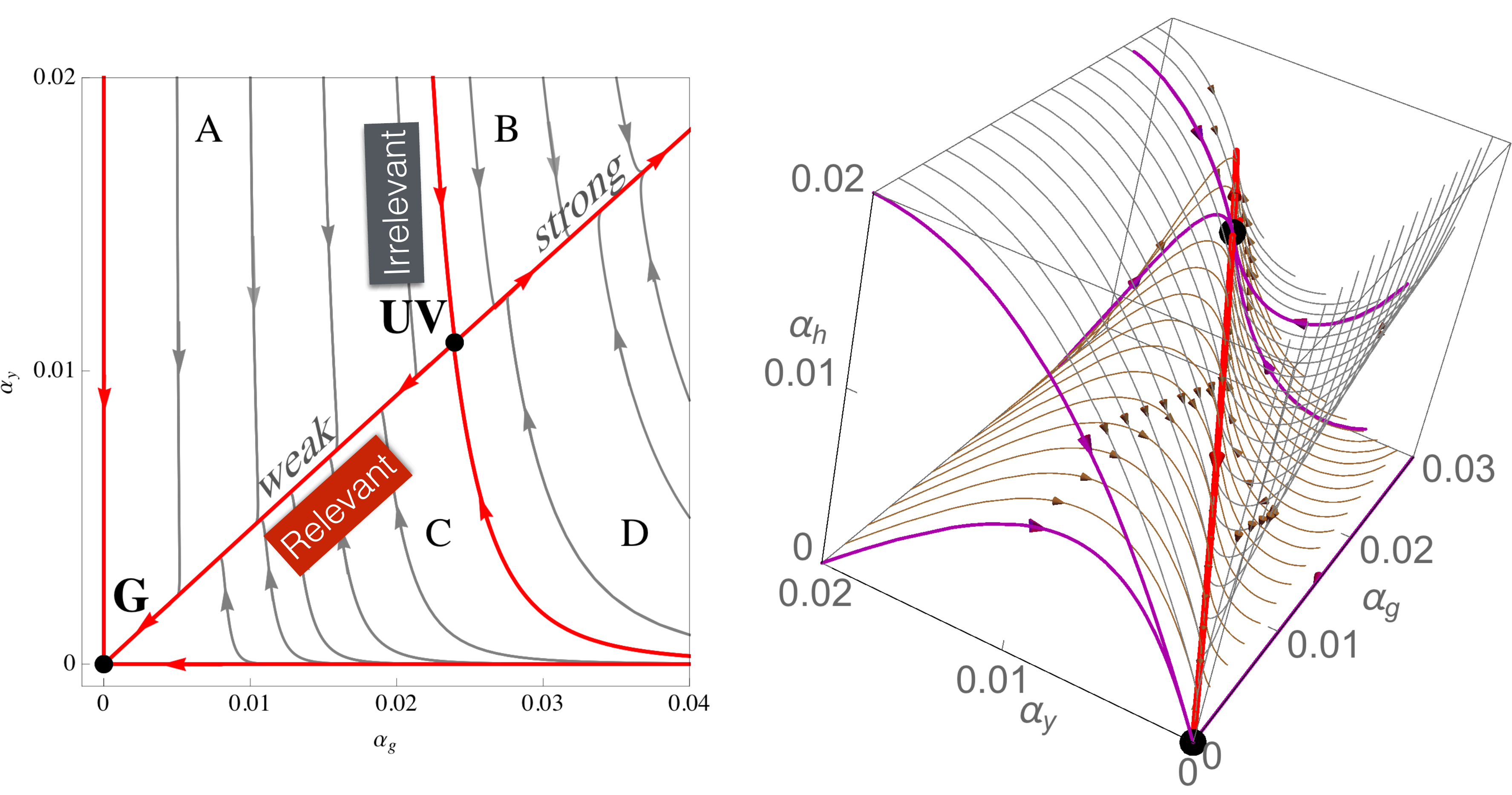}
\caption{\label{pPD} Review of the phase diagram of Ref.\ \cite{Litim:2014uca,Litim:2015iea}. 
The gauge-Yukawa subsector of couplings $(\al g,\al y)$ is shown
to leading order in the left-hand panel, while the gauge-Yukawa-scalar subsector $(\al g,\al y,\al h)$ at 
next-to-next-to-leading order in the right-hand panel for 
$\eps=0.05$ (corresponding to $N_f = 6$). Also shown are the ultraviolet and infrared fixed points (solid dots), the ultraviolet-safe trajectories 
(thick red line). A few trajectories are highlighted as thin magenta lines, and a few  
generic trajectories are shown as thin gray lines. Arrows point towards the infrared. 
For further details, we refer to Ref.\ \cite{Litim:2014uca,Litim:2015iea}.}
\end{figure}

The phase diagram of the theory was established in Ref.\ \cite{Litim:2014uca} at next-to-next-leading order accuracy and extended to the next-to-next-to-leading order in Ref.\ \cite{Litim:2015iea}, where the effects from the running scalar couplings were considered. For the sake of being self-contained, we summarise in Fig.\ \ref{pPD} the phase diagram of the theory shown in Ref.\ \cite{Litim:2015iea}. In the left panel, we show the renormalization group flow trajectories for the $(\al g,\al y)$ couplings,
while the right-hand panel illustrates the three-dimensional renormalization group flow that includes also the coupling $\al h$. The two plots include both the ultraviolet and infrared fixed points. We have also indicated in the left panel the relevant and irrelevant directions dictated by the signs of the scaling exponents. 

There are several nice and distinctive features of the analytically controllable and completely asymptotically safe dynamics summarised above. Collecting the features {\bf Safe~1} through {\bf Safe~5}, we find that the safe template considered above constitutes an ideal ground for investigating properties of the safe theories that for certain aspects resembles ${\cal N}=4$ super Yang-Mills theory. One of the similarities is the fact that along the line of physics all couplings are related. This is also a fundamental feature of the ${\cal N} =4$ super Yang-Mills theory due, however, to the high degree of supersymmetry. In the non-supersymmetric case the relations among the couplings are dynamical in nature being dictated by the dimension of the critical surface. One may access the non-perturbative regime of the theory by dialing $\epsilon$ large. We expect a critical value of $\epsilon$ above which the safety is lost. 

There is more physics to explore out of the safe gauge theory. In the phase diagram, emanating from the ultraviolet fixed point (marked UV in Figure 4), there are two relevant flows to the infrared, marked as weak and strong, respectively, in Figure 4. So far, we focused on the weak coupling phase of the flows. The strong coupling phase of the flow is also of interest, since the safe gauge theory in this region (which we label as {\it safe-QCD} hereafter) exhibits infrared dynamics reminiscent of QCD \footnote{This {\it safe-QCD} scenario is complementary to the scenario in which one adds novel matter and interactions that modify the ultraviolet dynamics of QCD to allow for an interacting fixed point at short distances. This last scenario was first proposed in \cite{Sannino:2015sel} and then realised in \cite{Abel:2017ujy,Abel:2017rwl,Abel:2018fls}.   }. 

In this phase, the `t Hooft coupling flows out of the ultraviolet fixed point at a finite value (of order the Veneziano ratio $\epsilon$) but eventually runs to infinite coupling in the infrared. This follows from the perturbative beta function shown in the left panel of Figure~\ref{running} (along the line of physics). We deduce that the `t Hooft coupling runs in the ultraviolet towards the interacting fixed point following a power law rather than displaying the asymptotically free logarithmic behavior. However it runs similarly to QCD in the infrared. 

We  therefore expect  that the {\it safe-QCD} theory develops spontaneous chiral symmetry breaking associated to the flavor symmetry  $G_f = SU(N_f)_L \times SU(N_f)_R$ and also confines the color charges. We know from QCD that both phenomena generate dynamical mass scales. In fact, this strong coupling phase is the non-abelian realization of the scenario put forward by Bardeen, Leung and Love \cite{Leung:1985sn}. Their idea was that, while the $G_c = U(1)$ gauge theory with $N_f$ massless fermions (QED) suffers of the zero charge problem or the triviality problem related to the existence of a Landau pole, once the theory is deformed by the introduction of four-fermion interactions, the resulting dynamics could develop a stable non-Gaussian ultraviolet fixed point. These authors noted that such deformation leads to the emergence of two physical phases. In one phase, the gauge coupling vanishes toward the infrared and  chiral symmetry remains intact. In the other phase, the gauge coupling grows toward the infrared and chiral symmetry is dynamically broken. With a one-dimensional critical surface (defining the line of physics), the flows out of the ultraviolet fixed point of our safe theory agree with the ones envisioned by Bardeen, Leung and Love. In our template, QED is replaced by its non-abelian version and the four-fermi interaction is replaced by the Yukawa coupling whose scalar field is dynamical and self-interacts quartically. The safe template is theoretically more manageable because, in the `t Hooft-Veneziano limit, the Yukawa and scalar couplings trigger the ultraviolet fixed point to emerge and drive its position within a controllable, weak coupling regime. 

While {\it safe QCD} behaves in the infrared in a manner that mimics QCD, there are several important differences. Firstly, the number of fermions (the size of the flavor group) is larger than the critical value for which asymptotic freedom is lost in the asymptotically free QCD. Secondly, in safe QCD, the fermions interact not only through gauge interactions but also through scalar interactions. Thirdly, the safe QCD, the number of fermions are larger than the critical number for the asymptotic freedom. Yet another, the scaling violations at short distances are not of the Bjorken type (as in asymptotically free QCD) but of the Kogut-Susskind type. 

Putting together, we claim that the strongly coupled region of {\it safe QCD} behaves as a non-abelian interacting plasma of massless fermions, scalars and gauge bosons in the ultraviolet (where the coupling is perturbatively weak)  and as a confined theory of hadrons with massive fermions, stemming from dynamical chiral symmetry breaking, at strong coupling in the infrared.  

\subsection{Supersymmetric Safe Templates}
\label{SafeSUSY}
Let us summarise the status of supersymmetric safe gauge theories. A safe theory (more precisely a weakly safe theory in the classification of Section 2.2) out of the ultraviolet fixed point can, in principle, flow to either a Gaussian infrared fixed point or to an interacting infrared fixed point. Above in Section 2.1, we discussed the first type of the renormalization group flow because it is both theoretically and phenomenologically important to assess whether non-asymptotically free theories can be ultraviolet complete, modulo the quantum gravity. As it should be clear by now, we just provided an affirmative answer to gauge-Yukawa theories that are remarkably similar in structure to the Standard Model in~\cite{Litim:2014uca} \footnote{To establish the existence of the safe theory nonperturbatively,  lattice simulation of the safe theory and continuum limit at the ultraviolet fixed point are imperative. For the pioneering work in this direction, see \cite{Leino:2019qwk}.}. 

Generalising and improving previous results of~\cite{Martin:2000cr}, the general conditions that must be (non-perturbatively) abided by non-asymptotically free supersymmetric theories to achieve safety  were put forward in~\cite{Intriligator:2015xxa}.  At least one chiral superfield must achieve a  large $R$ charge at the safe fixed point to ensure that the variation of the a-function between the safe and gaussian fixed points is positive as better elucidated in~\cite{Bajc:2016efj}. Models of this type were shown to exist in~\cite{Martin:2000cr,Bajc:2016efj,Bajc:2017xwx}. Another  possible way to elude the constraints~\cite{Intriligator:2015xxa,Bajc:2016efj} is to consider ultraviolet fixed points flowing to IR interacting fixed points. Within perturbation theory non-supersymmetric theories of this type were discovered in ~\cite{Esbensen:2015cjw}  and for supersymmetric theories in~\cite{Bond:2017suy,Bajc:2017xwx}.

Summarising the status for supersymmetric safety, we can say that theories abiding the constraints of~\cite{Intriligator:2015xxa} exist. Nevertheless specialising~\cite{Intriligator:2015xxa} to the case in which asymptotic freedom is lost ($N_f \geq 3N_c$) for super QCD with(out)
a meson~\cite{Martin:2000cr,Intriligator:2015xxa} one can show that these theories are unsafe for any number of matter fields. This is in agreement with the  $1/N_f$ studies of the supersymmetric beta
functions~\cite{Ferreira:1997bi,Ferreira:1997bi,Martin:2000cr} as explained in~\cite{Dondi:2019ivp}. The lack of safety of super QCD with(out) a meson should be contrasted with QCD at large number of flavors for which safety may be possible~\cite{Antipin:2017ebo} and QCD with a meson for which safety is a fact~\cite{Litim:2014uca}.

\section{Holography and Safe String Theory}
\label{label}

A safe theory asymptotes in the ultraviolet to a conformal field theory defined by an interacting fixed point, so it is natural to expect a holographically dual string theory. In this section, we shall work this out for the safe template of Section 2 in parallel to the ${\cal N}=4$ super Yang-Mills theory. The safe template of Section 2 was defined in the `t Hooft-Veneziano limit. Specifically, the way we formulated the template was within the perturbative regime of the Veneziano ratio $\epsilon$, $0 < \epsilon \ll 1$ such that the running couplings excurse within  ${\cal O}(\epsilon)$. According to the dictionary of the AdS/CFT correspondence, this would be in the regime where the string theory dual to the template lives in a highly curved background (though it asymptotically approaches the anti-de Sitter space). To effectively establish the safe gauge-string correspondence, we add to the safe template the ${\cal N}=4$ super Yang-Mills theory. We argued previously that this theory constitutes the simplest safe theory (as the beta function vanishes for any value of the coupling constant). 

\subsection{AdS/CFT Correspondence Revisited}
We recall that the AdS/CFT correspondence is definable for theories with a well-defined semiclassical limit. In the anti-de Sitter critical string theory, the limit is provided by weakly interacting strings. In the conformal field theory, it is provided by the planar fishnet expansion of Feynman diagrams. To pave our path toward the holography of safe gauge theories, we first recall the rationale behind for the correspondence between critical string theory in $AdS_5 \times S^5$ space and four-dimensional ${\cal N}=4$ super Yang-Mills theory. In the four-dimensional ${\cal N}=4$ super Yang-Mills theory, `t Hooft's large $N_c$ limit \cite{tHooft:1974pnl} reorganizes the gauge interactions in power expansion of $1/N_c^2$ of punctured Feynman diagrams \footnote{In all known conformal field theories that admit a holographic dual, it still is either the `t Hooft loop expansion parameter or the like.}. In the $AdS_5 \times S^5$ critical string theory, the weak coupling limit organizes the string interaction in the genus expansion of the punctured Riemann surfaces.

The four-dimensional ${\cal N}=4$ super Yang-Mills theory has an additional parameter; suitably rescaled, it is the `t Hooft coupling $\lambda = g^2 N$. Therefore, in the semiclassical limit, a physical observable ${\cal A}(N_c, g^2)$ of the Yang-Mills theory is organized as 
\bea
{\cal A}(N_c, g^2) = \sum_{h=0}^\infty \left({1 \over N_c^2} \right)^{-1 + h} A_h (\lambda) + \mbox{(nonperturbative)}\ ,
\label{thooftexpansion}
\eea
Here, $h$ denotes the genus of Riemann surface formed by a fishnet of Feynman diagrams with punctures specified by the Yang-Mills physical observables under consideration. The genus-$h$ contribution $A_h$ sums up all Yang-Mills interactions at a fixed genus $h$, so it depends on the `t Hooft coupling $\lambda$. For a fixed value of $\lambda$, the $N_c \rightarrow \infty$ limit is semiclassical, and so the Yang-Mills topological expansion Eq.(\ref{thooftexpansion})  is asymptotic with vanishing radius of convergence. 

The `t Hooft coupling parameter $\lambda$ takes an arbitrary non-negative fixed value, irrespective of the scale the dynamics is probed. So, the genus-$h$ contribution $A_h(\lambda)$ may be further expanded. This allows one to define the ${\cal N}=4$ super Yang-Mills theory in perturbation expansions of both $1/N_c^2$ and $\lambda$. For physical observables, one obtains
\begin{equation}
{\cal A} (N_c, g^2) \Big\vert_{\rm weak} = \sum_{h=0}^\infty \left({1 \over N_c^2} \right)^{-1 + h} \, \sum_{n=0}^\infty \lambda^n A_{h, n} + \mbox{(nonperturbative)} \ .
\label{yangmillsdoubleexpansion}
\end{equation}
At fixed genus $h$, the perturbation series in $\lambda$ is known to be convergent with a finite radius of convergence \cite{tHooft:1982uvh} (modulo issues related to renormalizations). For ${\cal N}=4$ super Yang-Mills theory, the radius of converges is $\pi$.  

The $AdS_5 \times S^5$ critical string theory depends on an additional parameter; suitably rescaled, it is the dimensionless curvature invariant $1/R^2$ measured in unit of the string scale $\ell_{\rm st}$. 
Therefore, in the semiclassical limit, a physical observable $\widetilde {\cal A}(g_{\rm st}, R^2)$ of the string theory is organized as 
\bea
\widetilde{A}(g_{\rm st}, R^2) = \sum_{h = 0}^\infty (g^2_{\rm st})^{-1 + h} \widetilde{A}_h (1/R^2) + \mbox{(nonperturbative)}. 
\label{stringexpansion} 
\eea
Here, $h$ denotes the genus of the Riemann surface formed by the string worldsheet with punctures specified by the string physical observables under consideration. The genus-$h$ contribution $\widetilde{A}_h$ sums up all gravitational curvature interactions, so it depends on the curvature invariant $1/R^2$.  For a fixed value of $1/R^2$, the $g_{\rm st} \rightarrow 0$ limit is semiclassical, and so the string topological expansion Eq.(\ref{stringexpansion}) is an asymptotic series with vanishing radius of convergence \cite{Gross:1988ib}. 

The (dimensionless) curvature invariant $1/R^2$ takes an arbitrary non-negative fixed value, irrespective the position the dynamics is probed. So, the genus-$h$ contribution $\widetilde{A}_h(1/R^2)$ may be further expanded. This allows one to define the $AdS_5 \times S^5$ string theory in perturbation expansions of both $g_{\rm st}^2$ and $1/R^2$. For physical observables, one obtains
\bea
\widetilde{A}(g_{\rm st}, R^2) \Big\vert_{\rm weak} = \sum_{h = 0}^\infty (g^2_{\rm st})^{-1 + h} \sum_{n}\left( {1 \over R^2} \right)^n \widetilde{A}_{h, n} + \mbox{(nonperturbative)}.
\label{stringdoubleexpansion}
\eea

The AdS/CFT correspondence is the statement that, for every physical observable, the Yang-Mills theory description given in Eq.(\ref{thooftexpansion}) and the string theory description of Eq.(\ref{stringexpansion}) are one and the same, in the sense that 
\bea
A_h(\lambda) = \widetilde{A}_h (1/R^2),
\label{adscft}
\eea
provided one identifies 
\bea
{1 \over N_c^2} = g_{\rm st}^2 \quad \mbox{and} \quad \lambda = R^4.  
\eea
With this identification, one finds from the defining double expansions of Eq.(\ref{yangmillsdoubleexpansion}) and Eq.(\ref{stringdoubleexpansion}), respectively, that the AdS/CFT correspondence is a claim for the strong-weak duality of $\lambda$ or $1/R^2$, respectively. Expressed in terms of Yang-Mills theory parameters, the physical observable Eq.(\ref{stringdoubleexpansion}) takes the form
\bea
\widetilde{A}(g_{\rm st}, R^2) = \sum_{h = 0}^\infty \left( {1 \over N_c^2} \right)^{-1 + h} \sum_{n}\left( {1 \over \sqrt{\lambda}} \right)^n \widetilde{A}_{h, n} + \mbox{(nonperturbative)}.
\label{stringdoubleexpansion}
\eea

The strong-weak duality is facilitated by the analyticity. As recalled above, the perturbative expansion of the genus-$h$ Yang-Mills contribution $A_{h}(\lambda)$ is convergent within a finite radius of convergence and it was within $|\lambda| \le \pi$ for ${\cal N}=4$ super Yang-Mills theory. By the analytic continuation in the complexified $\lambda$ plane, the contribution $A_h(\lambda)$ can be defined for arbitrarily large value of $\lambda$. We see that the perturbative series Eq.(\ref{stringdoubleexpansion}) for the genus-$h$ string contribution $\widetilde{A}_h$ 
ought to be an asymptotic expansion around $\lambda = \infty$. The non-analyticity of square-root branch cut around $\lambda = \infty$ was first discovered for the Wilson loop observable in \cite{Rey:1998ik, Maldacena:1998im, Rey:1998bq}, but it must occur for any physical observable. For Wilson loop observables, this was checked via explicit Feynman diagram computations in references \cite{Erickson:2000af} and  
\cite{Drukker:2000rr}. 

\subsection{Safe String Theory I: Weak Coupling Phase}
\label{3.2}
It should be possible to extend the AdS/CFT correspondence  above to theories that contain multiple parameters. This is  what we need for the safe gauge theories described above as they generically feature multiple coupling constants. Certainly, the safe template in Section 2 contained 6 parameters altogether; the number of colors $N_c$, the number of flavors $N_f$ for a given color, the Yukawa coupling $y$, the Higgs quartic couplings of single and double traces,  $h$ and  $v$, respectively. Both $N_c$ and $N_f$ can take arbitrarily large values, while the Higgs couplings are constrained by the classical stability of the ground state. By the AdS/CFT correspondence, our goal is to match the safe gauge theory defined at the ultraviolet fixed point with a string theory in a background that involves the anti-de Sitter space. We will do this by mimicking, as much as possible, the logical steps for the ${\cal N} = 4$ super Yang-Mills theory elucidated above. 

If the holographic picture holds, the dual string theory  must display a well-defined semiclassical limit, $g_{\rm st} \rightarrow 0$. We must then also have a single parameter in the safe gauge theory that facilitates the semiclassical limit. This immediately leads us to take the limit according to which both $N_c$ and $N_f$ approach infinity in a manner that keeps the ratio $N_f/N_c$ fixed. Only in this limit we achieve an $1/N_c^2$ counting. This is nothing but the 't Hooft-Veneziano limit we studied in Section 2, which intertwine the gauge symmetry $G_c$ with the flavor symmetry $G_f$. 
\begin{tcolorbox}
$\widetilde{\bf Safe~1}$:  To have a safe string theory dual, the flavor symmetry $G_f$ of the fermion and scalar fields is interlocked with the gauge symmetry $G_c$. This is facilitated by the  't Hooft-Veneziano limit. There may still survive an $N_c$ independent residual global symmetry $G_r$ in this limit. 
\end{tcolorbox}

With $\widetilde{\bf Safe~1}$, we take $1/N_c^2$ as the semiclassical expansion parameter. Any physical observable of the safe theory can be organized as
\bea
{\cal A}^{\rm Safe}(N_c, N_f, {\boldsymbol \alpha}) = \sum_{h=0}^\infty \left( {1 \over N_c^2}\right)^{-1 + h} A^{\rm Safe}_h (\epsilon, {\boldsymbol \alpha}) + \mbox{(nonperturbative)}.
\label{asexpansion}
\eea
Here, $\boldsymbol{\alpha}$ denotes the collection of other defining parameters of the safe gauge theory beyond  $N_c$ and $N_f$.  For our safe template these are  ${\boldsymbol \alpha} = (\alpha_g, \alpha_y, \alpha_h, \alpha_v)$ in Eq.(\ref{couplings}) and $\epsilon$ is the Veneziano ratio defined in Eq.(\ref{eps}). 
For fixed $\epsilon$ and $\boldsymbol \alpha$,  the $N_c \rightarrow \infty$ is a semiclassical limit, so the safe topological expansion is an asymptotic series of vanishing radius of convergence. 

As in the ${\cal N}=4$ super Yang-Mills theory, we may define the safe gauge theory as perturbative series in $\epsilon$ and ${\boldsymbol \alpha}$.  Any safe physical observable can then be expressed via the following multiple expansion
\bea
{\cal A}^{\rm Safe} (N_c, \epsilon, {\boldsymbol \alpha}) = \sum_{h=0}^\infty \left( {1 \over N_c^2}\right)^{-1 + h} 
\sum_{m = 0}^\infty \sum_{{\bf n} = 0}^\infty \cdots (\epsilon)^m (\alpha_1)^{n_1} (\alpha_2)^{n_2} \cdots A_{m, {\bf n}} + \mbox{(nonperturbative)}.
\nonumber\\
\label{asmultipleexpansion}
\eea

By the same reasoning as for the ${\cal N}=4$ super Yang-Mills theory, for a fixed genus $h$ and a fixed value of $\epsilon$, the series expansion in each of the parameters ${\boldsymbol \alpha}$ must have a finite radius convergence. On the other hand, at a fixed genus $h$ and a fixed value of $\boldsymbol \alpha$, the nature of the series expansion  in $\epsilon$ is not known. If the series were asymptotic, the safe gauge theory would contain two -- not one -- expansion parameters whose perturbative expansion is asymptotic. Such a theory would not admit a string theory (which has just one semiclassical limit of $g_{\rm st} \rightarrow 0$) as its holographic dual. 

The safe gauge theory  presented in Section 2  nicely gets around this issue. Recall that the safe template features the ultraviolet fixed point of the Banks-Zaks type, where the values of the couplings $\boldsymbol \alpha$, at the non-Gaussian fixed point, are now functions of the Veneziano ratio $\epsilon$. We conclude that
\begin{tcolorbox}
$\widetilde{\bf Safe~2}$: To have a safe string theory dual, the ratio $N_f / N_c$ ought to be intertwined to the values of the couplings $\boldsymbol \alpha$ at the non-Gaussian fixed point. 
\end{tcolorbox}

The asymptotic symmetry of the safe gauge theory at the ultraviolet fixed point constrains the structure of the multiple expansion Eq.(\ref{asmultipleexpansion}). It includes the $SO(4,2)$ conformal symmetry group and, if present, the $G_r$ global symmetry group. The coupling parameters $\epsilon$ and $\boldsymbol \alpha$ are all singlets under the $SO(3,1)$ Lorentz transformation and the $G_r$ global transformation. If the holography were to hold, there must be a massless scalar field dual to each of these parameters. By the $SO(4,2) \times G_r$ isometry, however, there is no such scalar field except the warp factor of the anti-de Sitter space. This means that the safe gauge theory must have the structure that all coupling parameters are proportional to one another near the ultraviolet fixed point. In particular, if the fixed point is at a finite coupling value, the scalar interactions should never hit the Landau pole. We thus have 
\begin{tcolorbox}
$\widetilde{\bf Safe~3}$: To have a safe string theory dual, the bulk scalar fields dual to the scalar quartic couplings must asymptotically range over a finite value. This property may hold without supersymmetry. 
\end{tcolorbox}
We already argued that the Veneziano ratio $\epsilon$ must be proportional to the coupling parameters, all through the renormalization group flow. This means that the critical surface near the ultraviolet critical point is 1-dimensional. 
\begin{tcolorbox}
$\widetilde{\bf Safe~4}$: To have a safe string theory dual, the couplings must flow over a finite but arbitrary interval set by the Veneziano ratio $\epsilon$ and the safe theories need to feature a one-dimensional ultraviolet critical surface.
\end{tcolorbox}

We already claimed that the ${\cal N}=4$ super Yang-Mills theory is the simplest safe gauge theory, whose gauge coupling does not run with energy scale. Its string theory dual lives in globally $AdS_5 \times S^5$ spacetime, where the curvature invariant $1/R^2$ is the same everywhere. For safe theories of the type presented in Section 2, the couplings run with the energy scale over a range set by the Veneziano ratio $\epsilon$. Accordingly, the string theory dual lives in asymptotically $AdS_5 \times X$ spacetime where the curvature invariant $1/R^2$ varies radially and asymptotes to a constant value set by the ultraviolet fixed point. Therefore, we have that
\begin{tcolorbox}
$\widetilde{\bf Safe~5}$: A holographic dual of the safe gauge theory is a string theory living in an asymptotically anti-de Sitter spacetime whose curvature invariant $1/R^2$ varies along the Liouville direction, forming a domain wall located at a critical scale. The infrared physics is encapsulated inside the wall while the ultraviolet physics is mapped to the region outside of the wall. 
\end{tcolorbox} 

We now demonstrate this. We expect that the holographic dual of the safe template in Section 2 is a 5-dimensional Type-0 non-critical string theory in the background of metric $g_{mn}$, dilaton $\phi$ of the NS-NS sector and volume-form $G_5$ of the R-R sector, all in the string frame \cite{Polyakov:1998ju}. The 4-dimensional Poincar\'e invariance puts the background to depend only on the Liouville direction \footnote{The non-critical string theory lives in the 5-dimensional spacetime but it may also live in an internal manifold  $X$ (of dimension less than 5) whose isometries encode the global symmetry $G_r$ of the safe gauge theory. Even in such situations, invariance under the spacetime Poincare and internal isometry transformations is sufficient to put the background in this form. }
\bea
&& ds^2_5 = g_{mn} d x^m d x^n = L_s^2(r) \left[ d r^2 + \eta_{\mu \nu} d x^\mu d x^\nu\right] \nonumber \\
&& \phi = \phi (r) \nonumber \\
&& G_5 = G(r) {N_c \over R_{\rm cr}} {\rm Vol}_5 \ . 
\nonumber
\eea
Here, $R_{\rm cr}$ is a reference length scale, which we set equal to the inverse of the critical transition scale $1/\Lambda_{\rm cr}$. We would like to see if a background $(g_{mn}, \phi, G_5)$ that encodes all the features of the safe theory can be constructed\footnote{A bottom-up approach of safe holography on a rigid AdS background was studied in \cite{Bajc:2019ari}.}.
 
The non-critical string theory of this sort was extensively studied in the context of holographic QCD \cite{Gursoy:2007cb,Gursoy:2007er}. In 5-dimensional spacetime, the 5-form $G_5$ is non-dynamical and its first integral defines a conserved charge, which can be identified with the number of color $N_c $ of the safe gauge theory. As argued above, a physical observable is evaluated by derivative expansion and the characteristic size is set by the reference length scale, $\partial_{\rm 5d} \simeq 1/ R_{\rm cr}$. In this expansion, the 5-form $G_5$ is the field originating from the R-R sector and so it contributes in powers of $||e^\phi G_5||^2$. In the `t Hooft limit $g_{\rm st} N_c$ fixed, this is $(g_{\rm st} N_c \ell_{\rm st} / R_{\rm cr})^2$ of order unity \footnote{In the $AdS_5 \times S^5$ supergravity dual to the ${\cal N}=4$ super Yang-Mills theory, $R_{\rm cr} = R_{\rm AdS}$ and this ratio is 1. The maximal supersymmetry then ensures that this ratio is exact and resummation is unnecessary. }, and so its contributions must be summed to all orders. A nontrivial dilaton potential would be generated from the resummation, but determining {\sl ab initio} its precise form is not feasible. Therefore, as in \cite{Gursoy:2007cb,Gursoy:2007er}, we will assume that, after a resummation of the 5-form $G_5$ contributions, the dynamics of the NS-NS sector is governed by a 5-dimensional dilaton gravity with a nontrivial dilaton potential, whose dynamics is described by the effective action
\bea
I_{\rm string} = M^3_{\rm st} \int d^5 x \sqrt{g_{\rm s}} e^{ - 2 \phi} \left[R_s (g) + 4 (\partial \phi)^2 + {1 \over R_{\rm cr}^2} {\cal V} (\phi) \right] \ .
\label{stringframe}
\eea
We will redefine the dilaton as $e^\Phi = N_c e^\phi$ and the metric in the Einstein frame as $g_e = e^{ - 4\Phi/3} g_s$, and study the on-shell dynamics of  
\bea
I_{\rm string} = M^3_{\rm st} N_c^2 \int d^5 x  \sqrt{g_e} \left[R_e (g) - {4 \over 3} (\partial \Phi)^2 + {1 \over R_{\rm cr}^2} V(\Phi) \right] \quad {\rm with } \quad   V(\Phi) = e^{\frac{4\phi}{3}} {\cal V}(\phi). \ \ \ \ \ 
\label{action}
\eea
We ask if we can construct an appropriate form of the dilaton potential $V(\Phi)$ that holographically encodes all the salient features of the safe gauge theory we collected in Section 2. 

For later convenience, we will introduce a prepotential (often called the superpotential of an underlying fake supersymmetry structure \cite{Cvetic:1991vp, Cvetic:1992bf, Cvetic:1992st, Skenderis:2006jq}) $W(\Phi)$ for the dilaton potential
\bea
V(\Phi) = -{4 \over 3} \left[ \left( {d W \over d \Phi} \right)^2 - \left( {4 \over 3} W \right)^2 \right].
\label{superpotential}
\eea
We also choose to foliate the warped 5-dimensional spacetime (in the Einstein frame) into  domain-wall coordinates \cite{Cvetic:1992bf, Cvetic:1992st, Ahn:1999dq}
and denote the derivative with respect to the $u$ coordinate as prime:
\bea
ds_5^2(g_e) = d u^2 + e^{2 A(u)} \eta_{\mu \nu} dx^\mu d x^\nu, \qquad {}' := {d \over  d u}.
\nonumber
\eea
Then, the on-shell configuration of the metric $g_e$ and the dilaton $\Phi$ is given by the first-order equations
\bea
\Phi'(u) &=& + {d W \over d \Phi} \, , \nonumber \\
A'(u) &=& - {4 \over 9} W.
\label{bps}
\eea

With the holographic prepotential $W$, we can readily encode the microscopic data of the safe gauge theory. As the dilaton $\Phi$ is the interpolating field of the `t Hooft coupling parameter and the warp factor 
$A$ is the logarithmic spacetime dilatation in the safe gauge theory, we holographically identify the beta function as
\bea
\beta (\lambda) \Big\vert_{\lambda = e^\Phi} = e^\Phi {d \Phi \over d A} = e^\Phi {\Phi'(u) \over A'(u)}.
\label{betafunction}
\eea
In globally anti-de Sitter space, $A(u) = - u$ (modulo constant shift) and $\beta (\lambda) = -  (e^\Phi)'$ matches with the definition of beta function. 

Identifying the safe gauge theory beta function with the holographic one Eq.(\ref{betafunction}), we can find the prepotential $W$. Explicitly, solving for $\beta(\lambda)/\lambda = - 9 /4 (d \log W/ d \Phi)$, we obtain
\bea
W(\Phi) = W_0 \exp \left( - {4 \over 9} \int^{e^\Phi} \! \! \! d \lambda {\beta (\lambda) \over \lambda^2} \right).
\label{superpotential-final}
\eea
The normalization factor $W_0$ is arbitrary and can be absorbed by shifting the integration domain; it should not affect physical observables. We can then obtain the profile of the dilaton $\Phi(u)$ by solving for the first equation of Eq.(\ref{bps}) and finally obtain the profile of the warp factor $A(u)$ by substituting $\Phi(u)$ into the second equation of Eq.(\ref{bps}). The upshot is that, given the beta function $\beta(\lambda)$ of the safe gauge theory as an input, we can holographically completely encode the renormalization group flow within the framework of Type-0 noncritical string theory.

We now solve for these profiles and demonstrate that the dilaton $\Phi$ background forms a domain-wall, interpolating between the two minima of $V(\Phi)$ representing the ultraviolet and the infrared fixed points and that the dilaton approaches these fixed points with the Kogut-Susskind power-law scaling. We start with the perturbative beta function of the safe gauge theory, shown in the left of Figure~\ref{pPD}
\bea
\beta (\lambda) = b \lambda (\lambda - \lambda_{\rm ir}) (\lambda_{\rm uv} - \lambda) \qquad (b > 0). 
\label{safebetafunction}
\eea
Here, for generality, we split the double zero at $\lambda = 0$ into two single zeros; physically, this can be done by defining the safe gauge theory in a spacetime away from 4 dimensions \cite{Codello:2016muj,Bajc:2019ari} or in certain supersymmetric safe theories \cite{Bond:2017suy,Bajc:2017xwx}. For the safe template in Section 2, $\lambda_{\rm uv} = {\cal O}(\epsilon)$ and $\lambda_{\rm ir} = 0$. We will explore this template by taking the limit $\lambda_{\rm ir} \rightarrow 0$ in the end. 

One might be concerned with the truncation of the safe beta function at two loop order. Moreover, the beta function is scheme-dependent starting from three loops.  We claim that the only physically relevant information is the existence of the fixed points as well as $C^1$ differentiability at the fixed points\footnote{The expansion of  physical quantities such as the mass anomalous dimension in the Veneziano parameter $\epsilon$ is scheme independent and converges rapidly \cite{Ryttov:2016hdp}.}. Additionally the domain-wall is a topological configuration and so it is stable against any continuous deformation of parameters in the on-shell dynamics.   

From the beta function, we find the prepotential $W(\lambda)$ by combining the two first-order equations, 
$\lambda d W / d \lambda = - (4/9) (\beta/\lambda) W$:
\bea
W (\lambda) 
&=& W_0 \exp \left[ - {4 \over 9} \int^\lambda {\beta (\tilde{\lambda}) \over \tilde{\lambda}^2} d \tilde{\lambda} \right]
\nonumber \\
&=& W_0 \exp \left[ {4 b \over 9} \left( \lambda_{\rm ir} \lambda_{\rm uv} \log \lambda - (\lambda_{\rm ir} + \lambda_{\rm uv} ) \lambda + {1 \over 2} \lambda^2 \right) \right] \ .  
\label{safeW}
\eea
Note that the prepotential $W$ is nowhere vanishing in the physical domain $\lambda \in (0, \infty)$. The exponential growth of $W$ for $\lambda \rightarrow \infty$ provides an infinite barrier and suppresses the dilaton $\Phi$ growing arbitrarily large, and so the truncation of the beta function at two-loop order is {\sl a posteriori} self-consistent (at least for the safe QCD, which we will explore further in the next subsection). 

The dilaton profile is given by the first-order equation
\bea
\int {d \lambda \over \beta (\lambda) W(\lambda)} \Big\vert_{\lambda = e^\Phi} = - {4 \over 9} u. 
\eea
As the prepotential $W$ is nowhere vanishing, the integral in the left-hand side is dominated by the zeros of the beta function. Moreover, $1/W$ furnishes exponential suppression of the integral for $\lambda> \lambda_{\rm uv}$ while $W$ vanishes as a power low for $\lambda$ below $\lambda_{\rm ir}$. Therefore, up to exponentially small corrections the dilaton profile, within the range $\lambda_{\rm ir},\lambda_{\rm uv}$, can be approximated by 
\bea
\int {d \lambda \over \beta (\lambda)} \Big\vert_{\lambda = e^\Phi} \simeq - {4 \over 9} u + {\cal O}(e^{- u}), 
\eea
 with the assumption that $\lambda$ takes a value over the interval $(\lambda_{\rm ir}, \lambda_{\rm uv})$. We plot the dilaton profile for two different sets of $\lambda_{\rm ir}, \lambda_{\rm uv}$ fixed points. 
\begin{figure}[htb]
\begin{center}
\includegraphics[width=.45\textwidth]{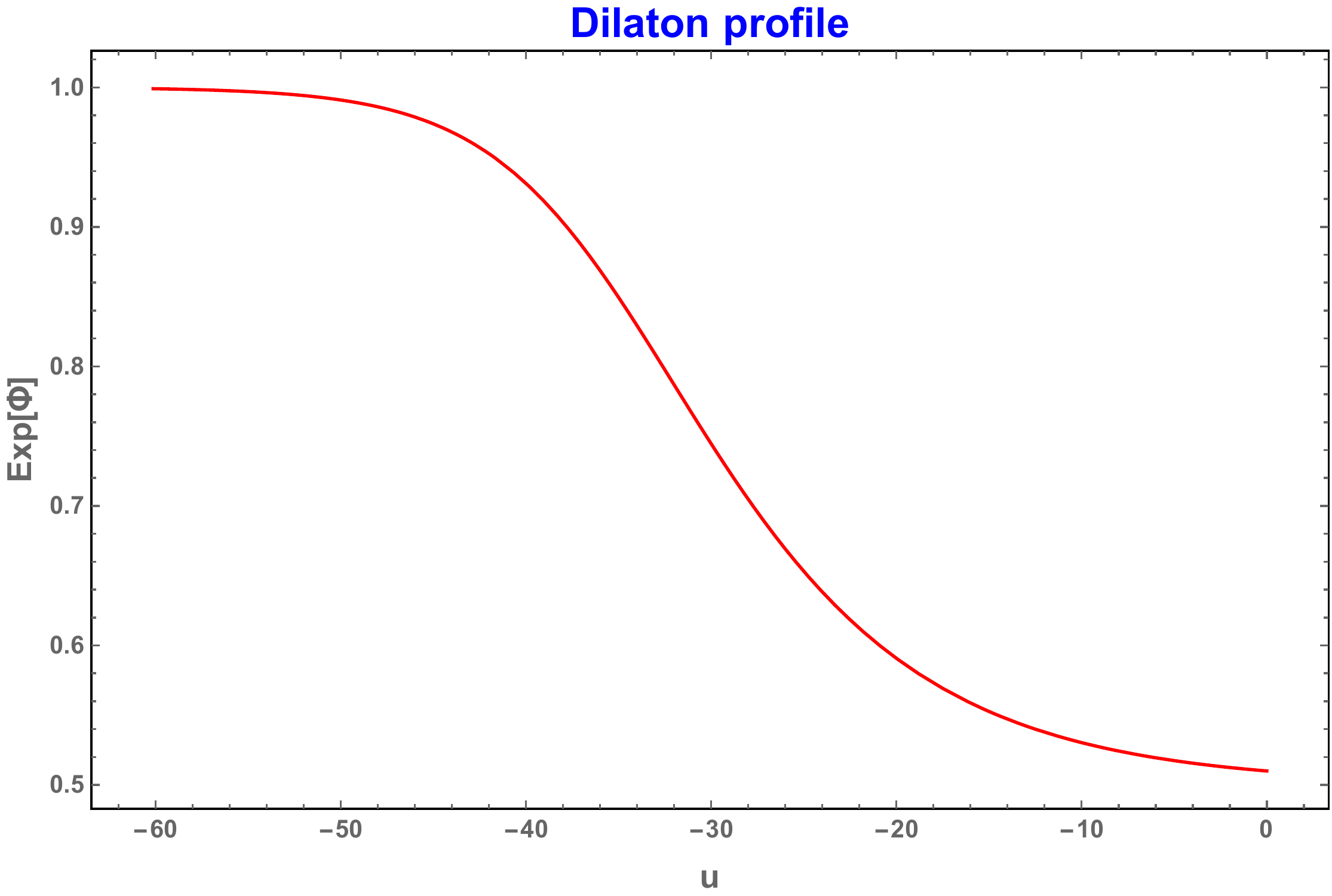} \hskip .4cm
\includegraphics[width=.45\textwidth]{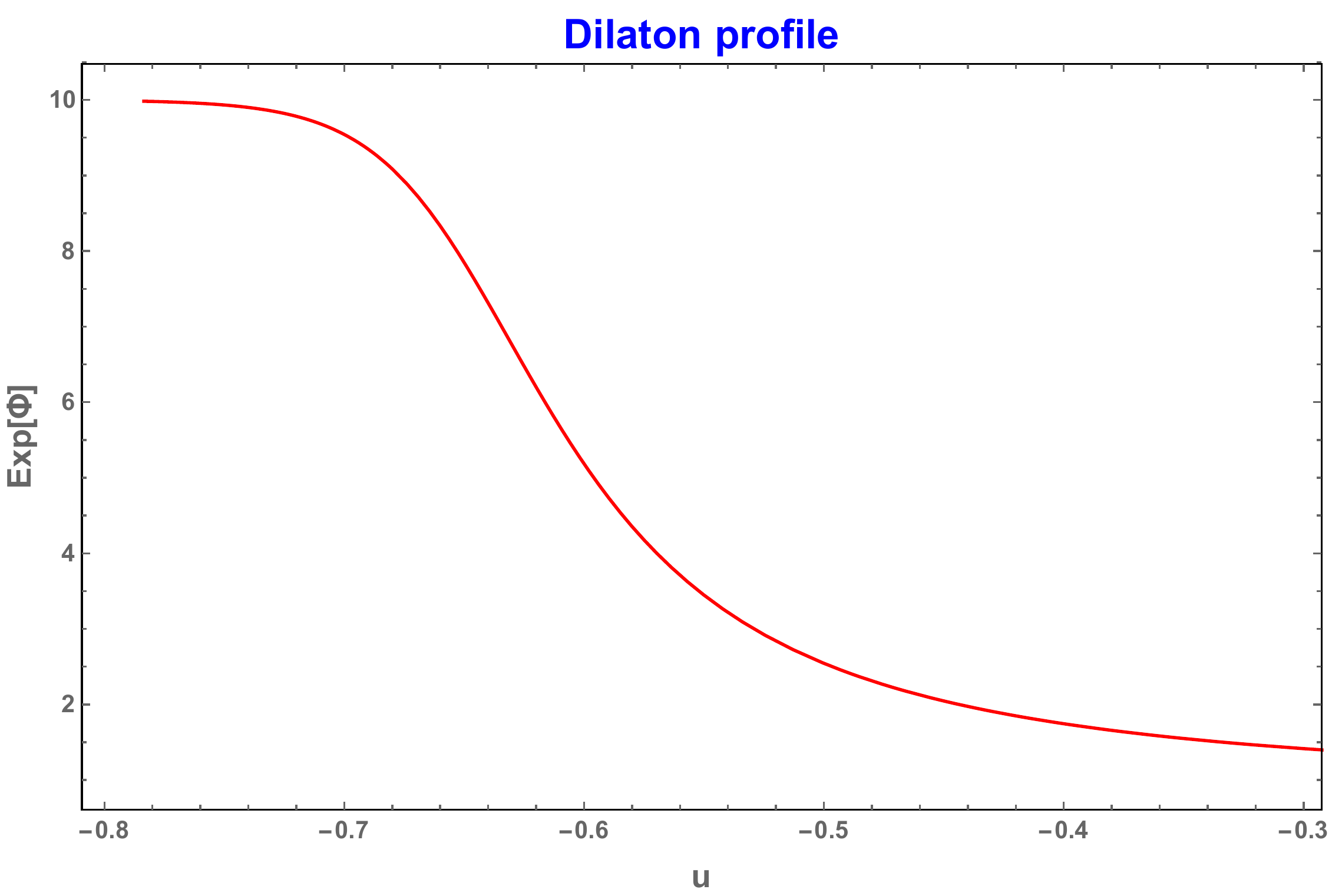} 
\caption{Profiles of the `t Hooft string coupling $e^\Phi$ plotted in $e^u$ coordinate. Left: weak coupling regime with $\lambda_{\rm ir} = 0.5, \lambda_{\rm uv} = 1$, Right: strong coupling regime with $\lambda_{\rm ir} = 1, \lambda_{\rm uv} = 10$. Note that abscissa scales differ by two orders of magnitude.}
\label{dilaton-domainwall}
\end{center}
\end{figure}
The plot clearly demonstrates the domain wall nature of the dilaton field. The domain wall is thick if the fixed points are at weak coupling regime and becomes thinner as the fixed points are pushed to strong coupling regime. 

To further show that the spacetime warp factor also forms a domain wall profile, we solve for the first-order equation for $A'(u) = - (4/9) W(\lambda(u))$ (assuming that $\Phi(u)$ is monotonic)
\bea
A(u) = - {4 \over 9} \int {W(\Phi) \over \left({d \Phi \over d u}\right) } d \Phi = { \int {W(\lambda) \over \beta(\lambda)  W(\lambda)  } } d \lambda = { + } \int {d \lambda \over \beta(\lambda)} \Big\vert_{u = u (\lambda)} = { -}{4 \over 9 } u.
\eea

The dilaton potential $V(\Phi)$ is also readily computed:
\bea
V(\Phi) &=& - {64 \over 27} \left[ \left( {\beta(\lambda) \over 3 \lambda} \right)^2 - 1 \right] |W|^2 \Big\vert_{\lambda = e^\Phi} \nonumber \\
&=& - {64 \over 27} \left[ {b^2 \over 9} (\lambda - \lambda_{\rm ir})^2 (\lambda - \lambda_{\rm uv})^2 - 1 \right] |W|^2 \Big\vert_{\lambda = e^\Phi}.
\eea
Recalling that the prepotential $W$ in Eq.(\ref{safeW}) exponentially grows away from $\lambda = (\lambda_{\rm ir} + \lambda_{\rm uv}) > \lambda_{\rm uv}$, we see that the Coleman-DeLuccia contribution $|W|^2$ of the dynamical gravity has the effect of lifting the otherwise degenerate extrema at $\lambda = \lambda_{uv}, \lambda_{ir}$ to $V (\lambda_{\rm uv}) < V(\lambda_{\rm ir})$.  

Summarizing, we explicitly checked $\widetilde{\bf Safe~5}$ by showing that the dilaton and the warp factor both exhibit the domain wall profile with asymptotic values set by the zeros of the prepotential $W(\Phi)$, equivalently, the two extrema of the dilaton potential $V(\Phi)$. 

\begin{tcolorbox}
$\widetilde{\bf Safe~6}$: In the safe string theory dual, the string coupling scales its asymptotic values according to power-laws. This is in part because the dilaton flows between two extrema of the potential and in part because the two extrema sit inside an infinite barrier. This power-law behavior is the holographic realization of the Kogut-Susskind scaling. 
\end{tcolorbox} 
\noindent
This is in sharp contrast to the holographic dual of QCD or asymptotically free gauge theories. In the latter cases, to encode the logarithmic behavior and so the Bjorken scaling, the dilaton potential ought to be bottomless so that the dilaton $\Phi$ indefinitely runs away to $-\infty$~\footnote{On the other hand, we will meet the similarity in the strong coupling phase in the next Section.}

Demonstrating this amounts to computing the scaling dimension $\Delta$ of the operator ${\cal O} = {1 \over 4} {\rm Tr} F_{\mu \nu} F^{\mu \nu}$ in the safe gauge theory. On general grounds, it is known to be \footnote{Note that the anomalous dimension is expressed in terms of $g$, not $g^2$. This error \cite{Gubser:2008yx} propagated boundlessly in holographic QCD literatures. This is not in contradiction with the fact that peturbation theory is in powers of $g^2$, simply because $\beta(g)$ is perturbatively given as odd powers in $g$ and $g (d/dg) = 2 g^2 (d/d g^2)$. }
\bea
\Delta_{F^2} = d + g {d \over d g} {\beta (g) \over g}.
\nonumber
\eea
In the safe gauge theory, the scaling dimension is extractable from two-point correlation function of the operator ${\cal O}$. In the holographic dual string theory, the scaling dimension is extractable from the fluctuation of the dilaton $\Phi$ and the warp factor $A$ in the domain-wall background. In the coordinate adopted, the field equations of both fields are
\bea
&& \Phi''(u) + 4 A'(u) \Phi'(u) {+ {3 \over 8} \frac{d V(\Phi)}{d\Phi}} = 0 \nonumber \\
&& A''(u) = -{4 \over 9} \Phi'^2(u) \nonumber \\
&& A'^2(u) = {1 \over 9} \Phi'^2(u) { + {1 \over 12} V(\Phi)}.
\eea
We expand this on-shell dynamics around either fixed points (where $\Phi = \Phi_*$ is constant-valued and $A = A_* = (4/9) u$) and linearize the fluctuation dynamics. From the two equations of the warp factor, we find that
\bea
&& \delta A'' = - {8 \over 9} \Phi'_* {(\delta \Phi)'} \nonumber \\
&& 2 A'_* (\delta A)' = {2 \over 9} \Phi_*' (\delta \Phi)' { + {1 \over 12} \frac{dV(\Phi)}{d\Phi} \Big\vert_{  \Phi =\Phi^\ast } \delta \Phi} \ . 
\eea
It follows that $(\delta A)''=0$ and the first-order correction of the warp factor is to rescale the background profile. 
From the fluctuation equation of the dilaton field, 
\bea
\delta \Phi'' (u) + 4 A'_* \delta \Phi' (u) ={- {3 \over 8} \frac{d^2V(\Phi)}{d\Phi^2}\Big\vert_{  \Phi =\Phi^\ast }  } \delta \Phi.
\eea
With the prepotential expression of the dilaton potential, we have 
\bea
m^2 _{\ast}= { + {3 \over 8} \frac{d^2V(\Phi)}{d\Phi^2}\Big\vert_{  \Phi =\Phi^\ast }  } = \left[ {16 \over 9} W - {d^2 W \over d \Phi^2} \right] {d^2 W \over d \Phi^2} \Big\vert_{\Phi_*}.
\eea
As $\beta (\lambda)$ and $\beta (\lambda) / \lambda$ vanish at either fixed points we have 
\begin{equation}
{\frac{d^2 W(\Phi_\ast)}{d\Phi^2 }\sim -(4/9) \frac{d\beta(\Phi_*)}{d\Phi} W(\Phi_*)}  \ ,\nonumber
\end{equation}
which vanishes at the infrared Gaussian fixed point and behaves as ${\cal O}(\epsilon^2)$ at the ultraviolet fixed point. Therefore, $(d^2W(\Phi_\ast)/d\Phi^2)^2$ either vanishes or it is at least of order $\epsilon^4$, respectively, and therefore subdominant to $W(\Phi_*) d^2W(\Phi_*)/d\Phi^2$. We conclude that $m^2_*$ of the dilaton fluctuation scales as $- \epsilon^2$: 
\bea
e^{\delta \Phi}  = \left[1 -  a \left( {r_c \over r}\right)^{ - b \epsilon^2} + \cdots \right], \qquad (a, b >0).
\label{warpfactor}
\eea
and that the scaling dimension $\Delta({\cal O})$ goes as $\Delta ({\cal O}) \simeq 4 + {1 \over 2} m_*^2 R_*^2$. 
Here, $b, c$ are non-universal, model-specific numerical factors. They both take positive values, since by $\widetilde{\rm Safe~3}$ it takes a finite value at the infinity $u \rightarrow \pm \infty$. We see that the power-law scaling is the holographic dual of the Kogut-Susskind scaling. 

We note that the salient features of the safe string theory $\widetilde{\bf Safe~1}$ through $\widetilde{\bf Safe~6}$ in this Section precisely match with the salient features of the safe gauge theory {\bf Safe~1} through {\bf Safe~6} in  Section~\ref{Safe}. We take this as convincing evidence for the existence of holographic duals of safe theories. Though the data points of the safe gauge theory in Section~\ref{Safe} were limited to weak coupling regime they can be extended to the non-perturbative regime for large 't-Hooft-Veneziano parameter $\epsilon$, additionally we stress that ${\cal N}=4$ super Yang-Mills theory (which is the simplest safe gauge theory) \footnote{That the ${\cal N}=4$ super Yang-Mills theory in the ultraviolet can flow to less supersymmetric conformal field theory in the infrared \cite{Girardello:1998pd} furnishes further credibility to this viewpoint. } and varieties of interacting conformal field theories with relevant flow to the infrared add data points for the non-perturbative regime. The geometry of  the holographic dual is an interpolating domain wall between two anti-de Sitter spaces of different scalar curvature. The global anti-de Sitter space, dual to the ${\cal N}=4$ super Yang-Mills theory, is the limit in which the domain wall flattens out. Additionally, varieties of known conformal field theories with a built-in `t Hooft-Veneziano limit may provide additional (albeit isolated) data points of non-perturbative safe gauge theories. We thus conjecture that
\vskip0.3cm
\begin{tcolorbox}
{\bf Conjecture: Safe Holography} \hfill\break
A safe $d$-dimensional gauge theory that admit `t Hooft-Veneziano limits is holographically dual to a $(d+1)$-dimensional safe noncritical string theory. The background of the holographic dual is a gravitational domain-wall, interpolating between two asymptotically anti-de Sitter spaces of different radius of curvature. 
\end{tcolorbox}

\subsection{Safe String Theory II: Strong Coupling Phase}

In Section~\ref{2.2}, we argued that the safe template also has the strong coupling phase, which we referred to as the {\it safe QCD}. Here we construct its holographic dual. We claim that the dual is the safe noncritical string theory of Section~\ref{2.2}, except that the on-shell background of warp factor and dilaton field obey different boundary conditions. As the analysis is nearly identical to Section~\ref{3.2}, we shall be brief in details and focus on the relevant differences. 

For the safe template of Section~\ref{SafeT}, the ultraviolet fixed point is located at $\alpha_g^{\rm uv} = {\cal O}(\epsilon)$. We still can make perturbative analysis reliably above (but not too far) the fixed point. Along the strong coupling line of physics, the safe beta function is essentially the same as the beta function of asymptotically free QCD. Therefore, in the deep infrared, the dilaton and warp factor would grow indefinitely and the chiral symmetry breaking and the color confinement should be probed via suitable physical observables. 

What distinguishes the safe QCD from the asymptotically free QCD is the way the ultraviolet fixed point is approached. While the asymptotically free dilaton potential runs monotonic with no maximum anywhere near $\Phi \sim - \infty$, the safe dilaton potential has a local maximum and is well behaved everywhere. Many difficulties for constructing the holographic QCD occurred because of the pathological profile with logarithmic runaway of the dilaton field. The safe dilaton $\Phi$ starts off the ultraviolet fixed point and grows large. Close to the $\lambda \sim \lambda_{\rm uv}$, the dilaton exhibits the Kogut-Susskind scaling
\bea
e^{\delta \Phi}  = \left[1 +  a \left( {r_c \over r}\right)^{ - b \epsilon^2} + \cdots \right], \qquad (a, b >0).
\label{warpfactor}
\eea
but with opposite sign of the model-specific parameters $a$. Such scaling makes the partons in a safe gauge theory behave very differently from partons in an asymptotically free gauge theory. 

We claim that, as the dilaton and warp factor flow away from the $u \sim 0$ asymptotic anti-de Sitter space (the holographic dual of the ultraviolet fixed point), their profile would evolve practically in the same way as holographic QCD.  Intuitively, we can understand this from the behavior of the perturbative beta function of the safe gauge theory. Recall that the input for constructing the holographic dual background is the gauge theory beta function, see 
Eq.(\ref{betafunction}). Up to two loops (which are renormalization scheme independent), the beta function of the `t Hooft coupling $\lambda$ is given in  Eq.(\ref{safebetafunction}). See the left panel of Figure~\ref{running}. Recall that $\lambda_{\rm ir}, \lambda_{\rm uv}$ are infrared and ultraviolet fixed points of the `t Hooft gauge coupling $\lambda$, and the safe template of Section~\ref{SafeT} corresponds to the limit $\lambda_{\rm ir}$ sent to 0. The safe theory in the strong coupling phase is defined by taking $\lambda$ slightly larger than $\lambda_{\rm uv}$ and approaching it as the ultraviolet cutoff is removed. The running of the Safe QCD `t Hooft coupling, in the strong coupling phase, mimics the one of  the pure Yang-Mills theory or ordinary QCD with sufficiently small number of fermions. We can see this by shifting the `t Hooft coupling constant as $\lambda = \lambda_{\rm uv} + \hat{\lambda}$, where $\hat{\lambda}$ is a positive definite, shifted coupling constant. The safe beta function is then reduced to 
\bea
\beta (\hat{\lambda}) \equiv {d \hat{\lambda} (\mu) \over d \log \mu} = - b_o \hat{\lambda} - b_1 \hat{\lambda}^2 - b_2 \hat{\lambda}^3 \ , 
\eea
where (with overall minus sign pulled out) the coefficients
\bea
b_0 = b \lambda_{\rm uv} (\lambda_{\rm uv} - \lambda_{\rm ir}), 
\qquad
b_1 = b (2 \lambda_{\rm uv} - \lambda_{\rm ir}),
\qquad
b_2 = b 
\eea
are all positive. The signs of the beta function coefficients match the ones of the pure Yang-Mills theory or  QCD with sufficiently small number of fermions~\footnote{We remind the readers the 2-loop beta function of SU(3) QCD dimensionally continued to $d$ less than 4 dimensions is given by 
\bea
\beta(\alpha_s) = - {4- d \over 2} \alpha_s -{33 - 2 N_f \over 12 \pi} \alpha_s^2 - {153 - 19 N_f \over 24 \pi^2} \alpha_s^3.
\label{qcdbeta}
\eea
For $N_f = 0, 1, \cdots, 8$, with the overall minus sign pulled out, the beta function coefficients are all positive. 
}. 

As the beta functions of respectively safe and asymptotically free QCD are similar at low energies it is natural to expect that the gauge-string correspondence of both QCDs is also similar in so far as the infrared dynamics is concerned; as much as we have the chiral symmetry breaking and the confinement in asymptotically free QCD, so do we in safe QCD.

\section{Further Discussions}
\label{Conclusions}
In this paper, we developed a program that aims at establishing a correspondence between safe quantum field theories and safe string theories. In order to achieve this {\it safe} gauge-string correspondence we started with introducing a safe template made by a gauge-Yukawa theory  featuring fermion matter and scalar fields. We then argued for the existence of a $(d+1)$-dimensional safe noncritical string theory dual for a given $d$-dimensional safe gauge theory admitting the 't Hooft-Veneziano limit. The string theory dual leaves on an asymptotically anti-de Sitter space with nontrivial dilaton and warp factor flows. The safe gauge theory template leads to two infrared phases, one flowing to weak coupling and another to strong coupling, respectively. This feature is shared by a wider class of safe quantum field theories. We study the holographic dual for both phases. We further argued that the strong coupling phase, that we dubbed {\it safe QCD}, possesses a holographic dual that is better controlled than the one for ordinary asymptotically free QCD.

We pointed out evidences that the ${\cal N}=4$ super Yang-Mills theory and associated AdS/CFT correspondence are a limit situation of the safe gauge theories and their safe gauge-string correspondence; flow of the `t Hooft coupling is frozen to the value at the ultraviolet fixed point. We emphasized their importance in the space of safe quantum field theories, as the theory along its line of fixed points provide data points of strongly coupled safe theories that are solvable through the AdS/CFT correspondence. We also mention that our argument is not limited to ${\cal N}=4$ super Yang-Mills theory and all known conformal field theories of varying degrees of supersymmetry and spacetime dimensions are a limit situation of the safe gauge theories. We emphasize that this interpretation is irrespective of whether the conformal field theory is isolated or equipped with a conformal manifold. 

We emphasized that safe theories greatly expand the space of quantum field theories. The development is in its infancy and so it would be helpful to corner the theory into various analytically controllable regimes. Two corners merit special mention. 
\begin{itemize}
\item The first example concerns a large flavor charge limit of the safe template~\cite{Orlando:2019hte}. In this limit,  the flavor symmetry is spontaneously broken; the symmetry breaking pattern and the spectroscopy of ground state energy were understood. At low energy, the theory features a decoupled confining Yang--Mills theory and a characteristic spectrum of type~I (relativistic) and type~II (non-relativistic) Goldstone bosons. The knowledge of the spectrum in the broken phase and the perturbative analysis at the safe fixed point allowed to establish scaling  dimensions of operators in the broken phase as a triple expansion in the inverse of the charge, the inverse of $N_f$ and the gauge coupling constant at the ultraviolet fixed point. The results further unveiled a number of noteworthy properties of the low-energy spectrum, vacuum energy and conformal properties of the theory, including the derivation of a new consistency condition for the relative sizes of the couplings at the fixed point. This method is amenable to generalizations to other safe templates. 

\item The second example regards the relevant complementary limit of finite $N_c$ and large $N_f$ for gauge-fermion theories. We recall that these theories (such as QCD) cannot be perturbatively safe when $N_f$ is just above the critical value associated to the loss of asymptotic freedom \cite{Caswell:1974gg}.  The beta function at leading order in the $1/N_f$ expansion was computed and investigated in \cite{PalanquesMestre:1983zy,Gracey:1996he,Holdom:2010qs,Pica:2010xq}. To this order one observes tantalising hints of the  existence of an interacting ultraviolet fixed point \cite{Pica:2010xq,Holdom:2010qs,Antipin:2017ebo}.  Related physical properties, including the scaling dimensions near the leading $1/N_f$ fixed point, were studied in Ref.\ \cite{Litim:2014uca} from which one can derive the glueball anomalous dimension \cite{Ryttov:2019aux}.  However, because the fixed point occurs by a delicate cancellation that might not be supported in the full theory it is crucial to study the analytic properties of the theory \cite{Dondi:2019ivp} as well as sub-leading corrections in $1/N_f^2$. At large but finite $N_f$ lattice simulations are the only way to address the ultraviolet dynamics of these theories. Pioneering lattice simulations for SU(2) gauge theories with 24 and 48 Dirac fermions in the fundamental representation were performed in \cite{Leino:2019qwk} while earlier studied appeared in \cite{deForcrand:2012vh}. These lattice simulations  constitute also the stepping stone towards analyzing the safe template dynamics used for non-perturbative values of the Veneziano parameter $\epsilon$. This will allow, de facto, to test the safe gauge-string correspondence which is notoriously hard to do in the supersymmetric case because the lattice regularization explicitly breaks supersymmetry. 
\end{itemize}
These developments are  interesting also in view of our work. The large charge limit, for example, offers an exciting playground for developing the associated holographic dual within our proposed safe gauge-string correspondence. Further, the finite number of colors and large number of matter fields limit helps shaping the boundary of the safe quantum field theories that can be further explored via holography \cite{reysannino2}.  

Our template's renormalization group fixed point structure and phases thereof are general enough that our safe holographic constructions should work for other safe quantum field theories. Many of these safe theories might not be all gauge theories but still possess a large-$N_f$ limit of flavor or global symmetries. It would be interesting to study them in the future as well. 

\section*{Acknowledgement}
We thank P. Argyres, B. Bajc, A. Lugo, D. Orlando,  S. Reffert, A. Sagnotti and G. Veneziano for stimulating discussions and comments. The work of S-J.R. was supported in part by the National Research Foundation of Korea Grants 2005-0093843 and 2012K2A1A9055280. The work of F.S. was partially supported by the Danish National Research Foundation Grant DNRF:90. 
\bibliography{bibAS}
\bibliographystyle{JHEP-2-2}

\end{document}